\documentclass[aps,pra,twocolumn,,groupedaddress,showpacs,floatfix]{revtex4-1}

\usepackage{times,amssymb,amsmath}
\usepackage{array}
\usepackage{ifpdf}
\usepackage[varg]{txfonts} 
\usepackage{graphicx}
\usepackage{upgreek}
\renewcommand{\vec}[1]{\mathbf{#1}}
\newcommand{\gvec}[1]{\boldsymbol{#1}}
\mathchardef\mhyphen="2D
\usepackage{cleveref}
\begin{document}

\title{Semi-classical approximation for the second harmonic generation in nanoparticles}
\author{Y. Pavlyukh}
\email[]{yaroslav.pavlyukh@physik.uni-halle.de}
\affiliation{Institut f\"{u}r Physik, Martin-Luther-Universit\"{a}t
  Halle-Wittenberg, 06120 Halle, Germany}
\author{J. Berakdar}
\affiliation{Institut f\"{u}r Physik, Martin-Luther-Universit\"{a}t
  Halle-Wittenberg, 06120 Halle, Germany}
\author{W. H\"{u}bner}
\affiliation{Department of Physics and Research Center OPTIMAS,
University of Kaiserslautern, PO Box 3049, 67653 Kaiserslautern, Germany}
\date{\today}
\begin{abstract}
Second harmonic generation by spherical nanoparticles is a non-local optical process that
can also be viewed as the result of the non-linear response of the thin interface layer.
The classical electrodynamic description, based e.g. on the non-linear Mie theory, entails
the knowledge of the dielectric function and the surface non-linear optical
susceptibility, both quantities are usually assumed to be predetermined, for instance from
experiment. We propose here an approach based on the semi-classical approximation for the
quantum sum-over-states expression that allows to capture the second-order optical process
from first principles. A key input is the electronic density, which can be obtained from
effective single particle approaches such as the density-functional theory in the local
density implementation.  We show that the resulting integral equations can be solved very
efficiently rendering thus the treatment of macroscopic systems. As an illustration we
present numerical results for the magic Na$_{2869}^-$ cluster.
\end{abstract}
\maketitle
\section{Introduction}
For the description of a wide range of phenomena in physics and chemistry one is faced
with the question of how to predict in a reliable and system specific way the response to
external electromagnetic fields that impart energy $\omega$ and possibly momentum
$\mathbf{k}$ to the system~\cite{mahan_many-particle_2000}.  To name but few examples, we
mention here the response of nanoparticles to light in the
linear~\cite{halas_plasmons_2011} and the non-linear regimes~\cite{roke_nonlinear_2009}.
For electrons as a perturbation we refer to the overview
~\cite{garcia_de_abajo_optical_2010}.  A well-studied route to address these issues is the
concept of the dielectric response functions which is detailed in standard books such as
in ~\cite{mahan_many-particle_2000}, but also in recent reviews, e.g. in
Refs. ~\cite{halas_plasmons_2011,roke_nonlinear_2009,garcia_de_abajo_optical_2010}.

For extended bulk metals and metallic surfaces the dielectric response was extensively
studied (Ref.~\cite{liebsch_electronic_1997,andersen_substrate_2002} and references
therein). An extension to finite-size systems brings about a number of new aspects that
complicate the theory but at the same time offer new opportunities for the occurrence of
new phenomena. For example, it is known that the second harmonic generation (SHG) is
forbidden in systems with an inversion symmetry on the dipole approximation
level. Therefore {\"O}stling, Stampfli, and Bennemann~\cite{oestling_theory_1993-1} and
Dewitz, H{\"u}bner, and Bennemann~\cite{dewitz_theory_1996} proposed the anharmonic
oscillator model to describe the second-order non-linear response from spherical
systems. A further step was undertaken by Dadap \emph{et
al.}~\cite{dadap_second-harmonic_1999} who considered the small-particle limit ($ka<1$)
and described SHG as a mixture of dipole and quadrupole excitation processes ($a$ stands
for the system size).  It was also shown how to connect the model output to experimentally
observed quantities, i.e. the elements of the non-linear optical susceptibility tensor
$\chi^{(2)}(\omega)$. In this approach the inversion symmetry is not a necessary
assumption as it is assumed that the second harmonic generation originates from the thin
surface layer, where the symmetry is broken anyway. This idea can be extended in several
directions as was demonstrated in numerous works: Pavlyukh and
H{\"u}bner~\cite{pavlyukh_nonlinear_2004} developed the non-linear Mie theory valid for
particles of arbitrary sizes, de~Beer and Roke included the sum-frequency generation
mechanism into the considerations~\cite{de_beer_nonlinear_2009}, the cylindrical geometry
was treated by Dadap~\cite{dadap_optical_2008} and finally the theory for arbitrarily
shaped particles was developed by de~Beer, Roke, and Dadap~\cite{de_beer_theory_2011}.

All these theories based on classical electrodynamics rely on the knowledge of the
frequency-dependent dielectric function and the non-linear optical susceptibility
tensor. With the fabrication processes being perfected and the system sizes tending
smaller and smaller one may wonder to which extent quantum effects are important and
whether it is justified to use the same susceptibility tensor to describe semi-infinite
and finite size systems on the nanometer range. To shed light on these issues, it is
desirable to have a quantum theory for the non-linear response on the nanoscale. The fully
atomistic approach seems to be out of reach for present computers, as currently maximally
hundreds of atoms are possible to treat using quantum chemistry codes. Yet, the
outstanding question is, how important are the electronic correlation effects and is there
possibly a way to stay on the solid quantum theory basis while treating larger systems?

There is an affirmative answer to these questions as was demonstrated recently in the
linear optics case by Prodan and Nordlander~\cite{prodan_structural_2003}. They succeeded
to push the limits of the time-dependent density functional theory (TDDFT) to metallic
systems containing millions of atoms. But at the same time they demonstrated that for
these system sizes the semi-classical approach becomes very accurate. This is a marked
finding as it allows to replace the complicated sum-over-states quantum mechanical
expression~\cite{orr_perturbation_1971} for the polarizability tensor with a single
integral equation. Consequently, there is only one parameter in the theory: the ionic
density distribution. With some reasonable assumptions about the ionic density such as in
the jellium model (this assumption is reasonable even for molecular structures, as we have
shown recently~\cite{pavlyukh_angular_2009,pavlyukh_kohn-sham_2010} for fullerenes. The
usefulness of the jellium model was demonstrated by the pioneering works of Ekardt on
sodium clusters~\cite{ekardt_dynamical_1984,ekardt_size-dependent_1985} or of Puska and
Nieminen on C$_{60}$~\cite{puska_photoabsorption_1993}.) we can obtain the ground state
electronic density from the solution of the Kohn-Sham equations and express the response
function in its terms. The Drude dielectric function  follows automatically.

The semi-classical approximation is rooted in the work of Mukhopadhyay and
Lundqvist~\cite{mukhopadhyay_density_1975} who derived the corresponding integral equation
within linear response theory. Their theory was applied in numerous cases ranging from
plasmons in the jellium model to collective resonances in
C$_{60}$~\cite{vasvari_collective_1996} or carbon
nanotubes~\cite{vasvari_collective_1997}. The equation can also be derived starting from
the quantum mechanical sum-over-states expressions~\cite{pavlyukh_fast_2012} and using the
assumption that the frequency of the external field is large compared with the
single-particle gap ($\omega\gg E_g$)~\cite{ichikawa_theory_2011}.

One may wonder why this program was not implemented for nonlinear optical
processes. As a matter of the fact, already in 1973 Wang, Chen and
Bower~\cite{wang_second_1973} classically treated second harmonic generation from
alkali metals. A decade later Apell~\cite{apell_non-local_1983} derived an expression for
the second harmonic current in the form:
\begin{equation}
\vec j(\vec r,2\omega)=\upalpha\big[
f(\vec E\cdot\vec\nabla)\vec E+\upbeta(\vec \nabla f)E^2+\upgamma f(\vec \nabla\cdot\vec
E)\vec E\big],
\label{eq:apell}
\end{equation}
where for the unperturbed ground state electron density in the form $n(\vec r)=n_0f(\vec
r)$ the $\upalpha$ parameter is a function of $n_0$ and $\omega$, whereas $\upbeta$ and
$\upgamma$ are constants. The theory gained less attention than, for example, the
Mukhopadhyay and Lundqvist work because the connection to quantum mechanics was lost. Here
we mention nonetheless numerous works in the field extending over more than three decades:
the small homogeneous spherical particle limit~\cite{hua_theory_1986}, the Rayleigh-Gans
scattering approximation for a lattice of such particles~\cite{martorell_scattering_1997},
second harmonic generation by two-dimensional
particles~\cite{valencia_second-harmonic_2003}, or a more recent treatment of arbitrary
geometries~\cite{zeng_classical_2009-1}. Notice that the assumption of homogeneous
electron density distribution within the sample is an inevitable component of such
classical theories.

It is possible to revive the theory by noting that the electric field ($\vec E$) in
Eq.~\eqref{eq:apell} must also include the \emph{induced field}. This simple observation
immediately raises the level of the theory to the random phase approximation (RPA) or, if
we include electronic correlations, to TDDFT level. Our manuscript is, hence, a
formalization of this message.

To this end we first derive an expression analogous to~\eqref{eq:apell} starting from the
sum-over-state quantum mechanical formula for the non-linear optical
susceptibility~\cite{orr_perturbation_1971} and employing the high-frequency
approximation. Although quite technical, we believe that this derivation
(Appendix~\ref{sec:A}) has its own merits as it establishes the equivalence of the
hydrodynamic approximation of Apell and the high-frequency semi-classical expansion. It is
interesting to notice that the $1/\omega^4$ asymptotic behavior of this quantity is at
variance with the results of Scandolo and Bassani~\cite{scandolo_kramers-kronig_1995} who
predict a $1/\omega^6$ decay. This seems, however, to be a direct consequence of the
assumption of the all-dipole excitation mechanism that underly their study. It is possible
to prove from general principles that the inclusion of the quadrupole excitations leads to
the $1/\omega^4$ asymptotics~\cite{satitkovitchai_dynamic_2009}.

Our derivation raises the question of whether it is sufficient to know the unperturbed
ground state density to obtain the lowest order approximation for an arbitrary response
function. We recall that from the point of view of the diagrammatic perturbation
theory~\cite{ward_calculation_1965} SHG comprises three processes in which the $2\omega$
photon is emitted before, between or after the absorption of two
$\omega$-photons. Consequently, one might wonder if each diagram of this expansion can
also be expressed in terms of $n(\vec r)$. As we demonstrate below, the answer is
negative, one additionally needs the one-particle density matrix $\gamma(\vec r,\vec r')$
whose diagonal elements are given by $n(\vec r)$.  This comes not as a surprise if we
consider the analogy with the orbital-free kinetic density functional
theory~\cite{wang_orbital-free_2002} where this matrix enters the kinetic energy term.

The non-linear susceptibility relates the \emph{second-order induced density} to the
\emph{local electric field}. The latter is a microscopic quantity that can be connected to
the external field by knowing the \emph{linear response function}. In the linear case it
is a standard route to get the RPA dielectric response. In the non-linear case the
procedure is, probably, less known. Therefore, we follow here a very pedagogical treatment
of Liebsch and Schaich~\cite{liebsch_second-harmonic_1989}. In fact, they applied a trick
suggested by Zangwill and Soven~\cite{zangwill_density-functional_1980} to avoid the
summation over the infinite number of unoccupied states for the calculation of the
response functions. This allowed them to describe SHG at simple metal surfaces as
effectively one dimensional systems (it is basically the same approach that enabled Prodan
and Nordlander~\cite{prodan_structural_2003} to treat very large spherical systems).

The outline of this work is as follows: In Sec.~\ref{sec:inteq} we start with the most
general relation between the induced densities and the effective fields [Eq.~(2.5) of
Liebsch and Schaich, Phys. Rev. B {\bf 40}, 5401 (1989)] and formulate integral equations
for the induced density with the non-interacting response functions as kernels. In
Sec.~\ref{sec:sym} we discuss the case of a spherical symmetry and the simplifications it
implies for the numerics. Finally, the second harmonic response of the magic Na$_{2869}^-$
cluster is presented in Sec.~\ref{sec:comp} for the illustration of our theory. Based on
our recently developed method for the solution of integral equations of this type we are
able to drastically reduce the computational cost from $\mathcal{O}(N^3)$ to
$\mathcal{O}(N)$, where $N$ is the number of mesh points to represent the density.

We use atomic units, i.~e., $\hbar=e=m_e=4\pi\varepsilon_0=1$ throughout. Two appendices
contain mathematical details to make the exposition in
Secs.~\ref{sec:inteq}-\ref{sec:comp} self-contained.
\section{Integral equations\label{sec:inteq}}
Our theory can easily be extended to include electron correlation effects by using the
exchange correlation functional of DFT. Our formulation here is at the random phase
approximation level. Within this approximation the  \emph{density-density response}
function can be obtained as
\begin{multline}
\chi(\vec r,\vec r';\omega)=\chi^{(0)}(\vec r,\vec r';\omega)+\int d\vec r_1\!\!
\int d \vec r_2\, \chi^{(0)}(\vec r,\vec r_1;\omega)\\ \times v(\vec r_1-\vec r_2)\,
\chi(\vec r_2,\vec r';\omega),
\end{multline}
where $v(\vec r-\vec r_1)$ is the Coulomb potential and $\chi^{(0)}(\vec r,\vec
r';\omega)$ is the \emph{non-interacting density-density response} function (known as
Lindhard function for the homogeneous electron gas in three dimensions, 3D):
\begin{equation}\label{Eq:chi0}
 \begin{split}
  \chi^{(0)}(\vec{r}, \vec{r}'; \omega)=2\sum_{i,j}&\frac{f_i-f_j}{\omega+E_i-E_j+i\eta}\\
       &\quad\times\psi_i(\vec{r})\psi_j^*(\vec{r})\psi_j(\vec{r}')\psi_i^*(\vec{r}'),
 \end{split}
\end{equation}
where $f$ is the Fermi function and $i$, $j$ refer to the collections of quantum numbers that
uniquely characterize the electronic states of the system. The infinitesimally small positive
number $\eta$ shifts the poles from the real axis and ensures, thus, the causality of the
response function. In what follows we will assume it can be incorporated in the $\omega$
variable.

Let us consider the response of the system subject to the harmonic electric field
oscillating at the frequency $\omega$, i.e. $\varphi^{(0)}(\vec r;t)=\varphi^{(0)}(\vec
r)\, e^{-i\omega t}$. Then, the induced density which oscillates at the frequency of the
applied field is given by:
\begin{multline}
\delta n^{(1)}(\vec r)=\int \!d\vec r' \chi(\vec r,\vec r';\omega)\,
\varphi^{0}(\vec r')\\
=\int\!\! d\vec r' \chi^{(0)}(\vec r,\vec r';\omega)\,\varphi^{(1)}(\vec r'),
\label{eq:n1}
\end{multline}
where $\varphi^{(1)}(\vec r)$ is the induced local field oscillating at the fundamental
frequency and consisting of the external potential plus the Hartree potential corresponding to
the induced density:
\begin{equation}
\varphi^{(1)}(\vec r)=\varphi^{(0)}(\vec r)+\int \!\!d\vec r' v(\vec r-\vec r')\,
\delta n^{(1)}(\vec r').
\label{eq:v1}
\end{equation}
The induced density oscillating at the double frequency results from the non-linear process
described by the $\chi^{(0)}_2$ response function and from the linear response to the
local field $\varphi^{(2)}(\vec r)$ oscillating at $2\omega$:
\begin{multline}
\delta n^{(2)}(\vec r)=\int\!\! d\vec r'\! \int \!\!d\vec r'' \chi^{(0)}_2(\vec r;\vec r',\vec r'';\omega)\,
\varphi^{(1)}(\vec r')\,\varphi^{(1)}(\vec r'')\\
+\int d\vec r' \chi^{(0)}(\vec r,\vec r';\omega) \,\varphi^{(2)}(\vec r').
\label{eq:n2}
\end{multline}
Because there is no external field at $2\omega$ the local field at this frequency is given
by the Hartree potential:
\begin{equation}
\varphi^{(2)}(\vec r)=\int \!\!d\vec r' v(\vec r-\vec r')\,
\delta n^{(2)}(\vec r').
\label{eq:v2}
\end{equation}
\Cref{eq:n1,eq:v1} yield the integral equation for the linear density:
\begin{multline}
\delta n^{(1)}(\vec r)=\xi^{(1)}(\vec r)+\\ \int \!\!d\vec r'\!\int\!\! d\vec r''\,
\chi^{(0)}(\vec r,\vec r';\omega)\,v(\vec r'-\vec r'')\,\delta n^{(1)}(\vec r''),
\label{eq:i1}
\end{multline}
while \cref{eq:n2,eq:v2} result in the integral equation for the second harmonic density:
\begin{multline}
\delta n^{(2)}(\vec r)=\xi^{(2)}(\vec r)+\\ \int\!\!d\vec r'\!\int\!\!d \vec r''\,
\chi^{(0)}(\vec r,\vec r';\omega)\,v(\vec r'-\vec r'')\,\delta n^{(2)}(\vec r'').
\label{eq:i2}
\end{multline}
We defined the source terms following \cite{liebsch_second-harmonic_1989} as:
\begin{eqnarray}
\xi^{(1)}(\vec r)&=&\int \!\!d\vec r' \chi^{(0)}(\vec r,\vec r';\omega)\,
\varphi^{(0)}(\vec r'),\label{eq:xi1_def}\\
\xi^{(2)}(\vec r)&=&\int\!\! d\vec r'\!\int\!\! d\vec r''
\chi^{(0)}_2(\vec r;\vec r',\vec r'';\omega)\,\varphi^{(1)}(\vec r')\,
\varphi^{(1)}(\vec r'').\label{eq:xi2_def}
\end{eqnarray}
Eqs.~(\ref{eq:i1}, \ref{eq:xi1_def}) and eqs.~(\ref{eq:i2}, \ref{eq:xi2_def}) when coupled
with an appropriate approximation for the non-interacting response functions allow to
completely describe the linear and the second-harmonic response.

The semi-classical approximation for the $\xi^{(1)}(\vec r)$ has already been derived
previously~\cite{pavlyukh_fast_2012} with the result:
\begin{equation}
\xi^{(1)}(\vec r)=\frac1{\omega^2}\Big(\vec\nabla n(\vec r)\cdot\vec \nabla\varphi^{(0)}(\vec r)
-n(\vec r)\Delta\varphi^{(0)}(\vec r)\Big).\label{eq:xi1}
\end{equation}
In Appendix~A we derive the second-harmonic generation source term~\eqref{eq:phi4_fin}
that can be represented as:
\begin{multline}
\xi^{(2)}(\vec r)=\frac1{2\omega^4}\vec\nabla\cdot
\Bigg(\vec\nabla\varphi^{(1)}(\vec r)\, \Big[n(\vec r)\Delta \varphi^{(1)}(\vec r)
+\Big(\nabla\varphi^{(1)}(\vec r)\cdot\nabla n(\vec r)\Big)\Big]\\
+\frac14 n(\vec r)\vec\nabla\Big(\vec\nabla\varphi^{(1)}(\vec r)\Big)^2\Bigg).\label{eq:xi2}
\end{multline}
Although these expressions look rather complicated they can be further simplified in the
case of spherical symmetry.
\section{Spherical symmetry\label{sec:sym}}
For spherically symmetric systems (i.~e.~$n(\vec r)\equiv n(r)$) \cref{eq:i1,eq:i2}
simplify considerably when projected onto spherical harmonics. Since these equations have
the same functional form we introduce an index $i=0,\,1,\,2$ to label corresponding
quantities. We use the \emph{multipole expansion} of the fields:
\[
\varphi^{(i)}(\vec r)=\sum_{\ell m}\varphi^{(i)}_{\ell m}(r) Y_{\ell m}\left(\frac{\vec
r}{r}\right).
\]
This is a general form consistent with the spherical symmetry. Similar expansions will be
used for the densities and the source terms:
\[
\delta n^{(i)}(\vec r)=\sum_{\ell m}\delta n^{(i)}_{\ell m}(r) Y_{\ell
m}\left(\frac{\vec r}{r}\right),\quad
\xi^{(i)}(\vec r)=\frac{1}{\omega^{2i}}\sum_{\ell m}\xi^{(i)}_{\ell m}(r) Y_{\ell
m}\left(\frac{\vec r}{r}\right).
\]
For numerical calculations in this work we restrict ourselves to the mixture of the dipole
and quadrupole excitations ($\varphi^{(0)}_{\ell m}=0$ for $\ell>2$). Further
simplifications can be achieved by considering the external field to be a plane-wave:
\begin{equation}
\exp(i\vec k\vec r)=4\pi\sum_{\ell m} i^\ell
j_\ell(kr)Y_{\ell m}(\gvec{\Omega}_r)Y_{\ell m}(\gvec{\Omega}_k),
\label{eq:pw}
\end{equation}
and when we align the coordinate system along the $k$-direction:
\begin{equation}
\exp(i\vec k\vec r)\equiv e^{ikr\cos\theta}=4\pi\sum_\ell i^\ell \sqrt{\frac{2\ell+1}{4\pi}}
j_\ell(kr)Y_{\ell 0}\left(\frac{\vec r}{r}\right).
\label{eq:pw2}
\end{equation}
When the wave-number $k$ is small compared to the dimension of the system $a$
(i.~e. $ka<1$) it is justified to replace the \emph{spherical bessel functions} with their
small argument approximations $j_\ell(z)\sim z^n/(2\ell+1)!!$, use the definition of the
\emph{multipole moments}:
\[
Q_{\ell m}(\vec r)=\sqrt{\frac{4\pi}{2\ell+1}}r^\ell
Y_{\ell  m}\left(\frac{\vec r}{r}\right),
\]
and express the external potential as:
\begin{eqnarray}
\varphi^{(0)}(\vec r)&=&\sum_\ell\frac{(ik)^\ell}{(2\ell-1)!!}Q_{\ell 0}(\vec
r),\quad\text{or}\\
\varphi^{(0)}_{\ell
  m}(r)&=&\frac{(ik)^\ell}{(2\ell-1)!!}\sqrt{\frac{4\pi}{2\ell+1}}r^\ell\delta_{m0},
\label{eq:phi0_lm}
\end{eqnarray}
where each term is a harmonic function i.~e.~$\Delta Q_{\ell m}(\vec r)=0$.  Thus, in this
approximation, it is sufficient to consider only the first term in \eqref{eq:xi1}:
\begin{equation}
\xi^{(1)}(\vec r)=\frac1{\omega^2}\vec\nabla n(\vec r)
\cdot\vec \nabla\varphi^{(0)}(\vec r).\label{eq:xi1_fin}
\end{equation}
 With the help of \cref{eq:n1,eq:v1} we have:
\[
n\Delta \varphi^{(1)}=-\omega_p^2\,\delta n^{(1)},\quad
(\nabla\varphi^{(1)}\cdot\nabla n)=(\omega^2-\omega_p^2)\,\delta n^{(1)},
\]
where we introduced the \emph{local plasmon frequency} in analogy with the expression for
homogenous systems $\omega_p^2(r)=4\pi n(r)$. Finally, the second-order source term can be
given as a sum of two contributions $\xi^{(2)}(\vec r)=\xi^{(2a)}(\vec r)+\xi^{(2b)}(\vec
r)$:
\begin{subequations}\label{eq:xi2_fin}
\begin{eqnarray}
\xi^{(2a)}(\vec r)&=&\frac1{2\omega^4}\vec\nabla\cdot
\Bigg(\delta n^{(1)}(\vec r)\,\Big(\omega^2-2\omega_p^2(r)\Big)\,
\vec\nabla\varphi^{(1)}(\vec r)\Bigg),\label{eq:xi2a}\\
\xi^{(2b)}(\vec r)&=&\frac1{8\omega^4}\vec\nabla\cdot \Bigg(n(r)\vec\nabla\,
\Big(\vec\nabla\varphi^{(1)}(\vec r)\Big)^2\Bigg).\label{eq:xi2b}
\end{eqnarray}
\end{subequations}
The physical meaning of these two terms can be inferred by introducing the induced
electric field $\vec E=-\vec\nabla \varphi^{(1)}$ in e.g. Eq.~\eqref{eq:xi2}. The first
term contains a contribution from the electric quadrupole term $\vec E (\vec
\nabla\cdot\vec E)$ and from the density variation (the surface dipole term) $\vec E (\vec
E\cdot\vec \nabla n)$, whereas the second term upon the use of the vector identity
$\frac12\nabla E^2=\vec E\times\vec \nabla\times \vec E+(\vec E\cdot\vec \nabla)\vec E$
and one of the Maxwell equations $\vec E\times\vec \nabla\times \vec E=i\omega/c\vec
E\times\vec H$ can be interpreted as the magnetic ``bulk'' term and a surface
term. According to the analysis of Wang, Cheng and Bower~\cite{wang_second_1973} and
Apell~\cite{apell_non-local_1983} the contribution from $\xi^{(2b)}$ is small in the case
of the non-linear response from the surfaces and the incident electric field polarized in
the plane of incidence. These arguments loose their power in the case of the non-linear
response from spherical objects. Thus, both terms must be considered. We will show below
that the treatment of the second term is much more involved. Therefore, to illustrate our
approach only the first term will be included into the numerical algorithm.
\subsection{Computational scheme}
With these results in hands the numerical algorithm can be formulated as follows:
\begin{itemize}
\item[i)] The integral equation \eqref{eq:i1} is solved with a source term
  ~\eqref{eq:xi1_fin}. It yields the first order density $\delta n^{(1)}(\vec r)$ and,
  therefore, the local potential $\varphi^{(1)}(\vec r)$;
\item[ii)] Eq.~\eqref{eq:xi2_fin} is applied to generate the source term for the equation
  \eqref{eq:i2};
\item[iii)] This integral equation is solved similarly to \eqref{eq:i1} in order to obtain the
  second order density.
\end{itemize}
\subsection{Linear response} In this subsection we focus on the first point of the
  program. Already on this level our approach is valuable as it provides the optical
  absorption cross-section.
\paragraph{The linear source term:}
Using \eqref{eq:xi1_fin} and the explicit form of the potential \eqref{eq:phi0_lm} we
obtain:
\begin{equation}
\xi^{(1)}_{\ell m}(r)=\ell n'(r)\frac{\varphi^{(0)}_{\ell m}(r)}{r},
\label{eq:xi1_sph}
\end{equation}
where $n'(r)$ denotes the derivative of the equilibrium density function with respect to
the radial coordinate.
\paragraph{The integral kernel:}
The response functions are invariant under the rotations of the system as a whole. Thus,
in the linear case it can be expanded as:
\[
\chi(\vec r,\vec r';\omega)=
\sum_{l\,\mu}\chi_{l\,\mu}(r,r';\omega)Y^*_{l\,\mu}(\Omega)Y_{l\,\mu}(\Omega').
\]
Thus, the integral kernel of \cref{eq:i1,eq:i2} can be projected onto the angular momentum
eigenstates:
\begin{multline}
K_{\ell}(r,r')=\int\!d\Omega\, Y^*_{\ell m}(\Omega)\\
\times\int\!d\Omega'\,\vec\nabla n(r) \cdot \vec\nabla_{\vec r} v(|\vec r-\vec r'|)  Y_{\ell
  m}(\Omega').\label{eq:Oint}
\end{multline}
It can be integrated using the spherical harmonics expression of the Coulomb potential
  [see Sec.~3.6 of Jackson (Ref.~\onlinecite{jackson_classical_1999})]:
\[
v(\vec r-\vec r')=\frac{1}{|\vec r-\vec r'|}=\sum_l\sum_{\mu=-l}^{l}\frac{4\pi}{2l+1}
\frac{r_<^l}{r_>^{l+1}}Y^*_{l\mu}(\Omega)Y_{l\mu}(\Omega'),
\]
where $r_<$ ($r_>$) is the smaller (larger) of $r$ and $r'$. The gradient of the spherical
harmonics need not to be considered in view of the symmetry of the density, the angular
integration can be done beforehand and we obtain for~\eqref{eq:Oint}:
\begin{multline}
K_\ell(r,r')= \frac{4\pi}{2\ell+1}n'(r)\frac{\partial }{\partial
r}\left(\frac{r_<^\ell}{r_>^{\ell+1}}\right)\\
=-\frac{4\pi}{2\ell+1}n'(r)\frac{r^{\ell-1}}{r'^{\ell+1}}G^\text{sph}_\ell(r,r'),
\label{eq:asph}
\end{multline}
with the spherical $\ell$-pole Green function defined as
\begin{equation}
G^\text{sph}_\ell(r,r')=(\ell+1)\,\theta(r-r')
\left(\frac{r'}{r}\right)^{2\ell+1}\!\!\!-\ell\theta(r'-r).\label{eq:G_sph}
\end{equation}
With the help of \eqref{eq:asph} the integral equations \eqref{eq:i1} and \eqref{eq:i2}
can be written in the unified form:
\begin{multline}
\left(1-\frac{\omega_p^2(r)}{\omega^2}\right)\delta n^{(i)}_{\ell m}(r;\omega)
=\frac{\xi^{(i)}_\ell(r;\omega)}{\omega^{2i}}
-\frac{4\pi}{2\ell+1}\frac{n'(r)}{\omega^2}r^{\ell-1}\\
\times\int_0^\infty \!\!d r' \left(\frac{1}{r'}\right)^{\ell-1}\!G^\text{sph}_\ell(r,r')\,
\delta n^{(i)}_{\ell m}(r';\omega).\label{eq:int_nsph}
\end{multline}
This equation is central for our theory in both the linear and the non-linear cases. An
efficient method of its solution exists~\cite{pavlyukh_fast_2012}. In the linear case the
equation takes a more symmetric form when re-formulated in terms of the polarizability
function.
\paragraph{Observables:}
The $\ell$-\emph{pole frequency-dependent polarizability} defined as:
\[
\alpha_{\ell m}(\omega)=-\int\!\!d\vec r\int\!\!d \vec r'\,Q^*_{\ell m}(\vec r)
\chi(\vec r,\vec r';\omega)Q_{\ell m}(\vec r'),
\]
can be computed as the response to the field
$\phi_\ell^{(0)}(r)=\sqrt{\frac{4\pi}{2\ell+1}}r^\ell$. Let us introduce the
position-dependent polarizability as:
\[
\alpha_\ell(r;\omega)=\sqrt{\frac{4\pi}{2\ell+1}}\delta n^{(1)}_{\ell 0}(r;\omega)\,
\]
and use the explicit form \eqref{eq:xi1_sph} for the linear source. Substituting these
definitions in \eqref{eq:int_nsph} we obtain the following integral equation:
\begin{multline}
\alpha_\ell(r;\omega)=\alpha^{(0)}_\ell(r;\omega)\\
\times\Bigg[\ell-\int_0^\infty d r'
\left(\frac{1}{r'}\right)^{\ell-1}\!G^\text{sph}_\ell(r,r')\,
\alpha_\ell(r';\omega)\Bigg].\label{eq:int_asph}
\end{multline}
where
\[
\alpha^{(0)}_\ell(r;\omega)
=-\frac{4\pi}{2\ell+1}\frac{r^{\ell-1} n'(r)}{\omega^2-\omega^2_p(r)}.
\]
In the case of the dipolar response our result \eqref{eq:int_nsph} coincides with Eq.~(5)
of Prodan and Nordlander~(Ref.~\onlinecite{prodan_structural_2003}). Finally the
frequency-dependent polarizability is represented as the integral:
\begin{equation}
\alpha_{\ell 0}(\omega)=\int_0^\infty \!\!dr\,r^{\ell+2}\alpha^{(0)}_\ell(r;\omega).
\end{equation}
\subsection{The non-linear  source term}
The source term for the second-order response is considerably more complicated. The
central quantity is the induced potential $\varphi^{(1)}(\vec r)$. It can be evaluated by
the integration of \eqref{eq:v1}:
\begin{equation}
\varphi^{(1)}_{\ell m}(r)=\varphi^{(0)}_{\ell m}(r)+\frac{4\pi}{2\ell+1}\int\!\!dr'\,
r'^2\frac{r_<^\ell}{r_>^{\ell+1}}\delta n^{(1)}_{\ell m} (r'),
\end{equation}
where the density is obtained from the solution of \eqref{eq:int_nsph}. It is easy to see
that in accordance with the earlier assumption it is sufficient to restrict ourselves to
the cases of $\ell=1,\,2$ and $m=0$. The most difficult part of the derivation: the
transition from $\varphi^{(1)}(\vec r)$ to $\xi^{(2)}(\vec r)$ can probably be obtained in
closed form, however, for practical applications it is sufficient to have shorter form
small-$\ell$ solutions. They can be obtained with the {\sc maple} computer algebra
package:
\begin{multline}
\xi^{(2a)}_1=\frac{1}{2\sqrt{5\pi}}\Bigg[
u'\Big(n_1 f_2'+n_2f_1'\Big)+\frac{3u}{r^2}\Big(n_1f_2+n_2f_1\Big)\\
+u\,\Big(n_1'f_2'+n_2'f_1'-8\pi n_1n_2\Big)\Bigg],
\end{multline}
\begin{multline}
\xi^{(2a)}_2=\frac{1}{2\sqrt{5\pi}}\Bigg[ f_1'\Big(u'n_1
+un_1'\Big)-\frac{u}{r^2}\Big(n_1f_1+4\pi r^2 n_1^2\Big)\Bigg],
\end{multline}
where we introduced for brevity $\xi^{(2a)}_\ell\equiv\xi^{(2a)}_{\ell 0}(r)$, $u\equiv
\omega^2-2\omega^2_p(r)$, $n_\ell\equiv\delta n_{\ell 0}^{(1)}(r)$,
$f_\ell\equiv\varphi^{(1)}_{\ell 0}(r)$.

The second part $\xi^{(2b)}(\vec r) $ is presented in Appendix B for reference, however,
will not be included into the numerical algorithm.
\section{Numerical results\label{sec:comp}}
\begin{figure}[t!]
\includegraphics[width=0.7\columnwidth]{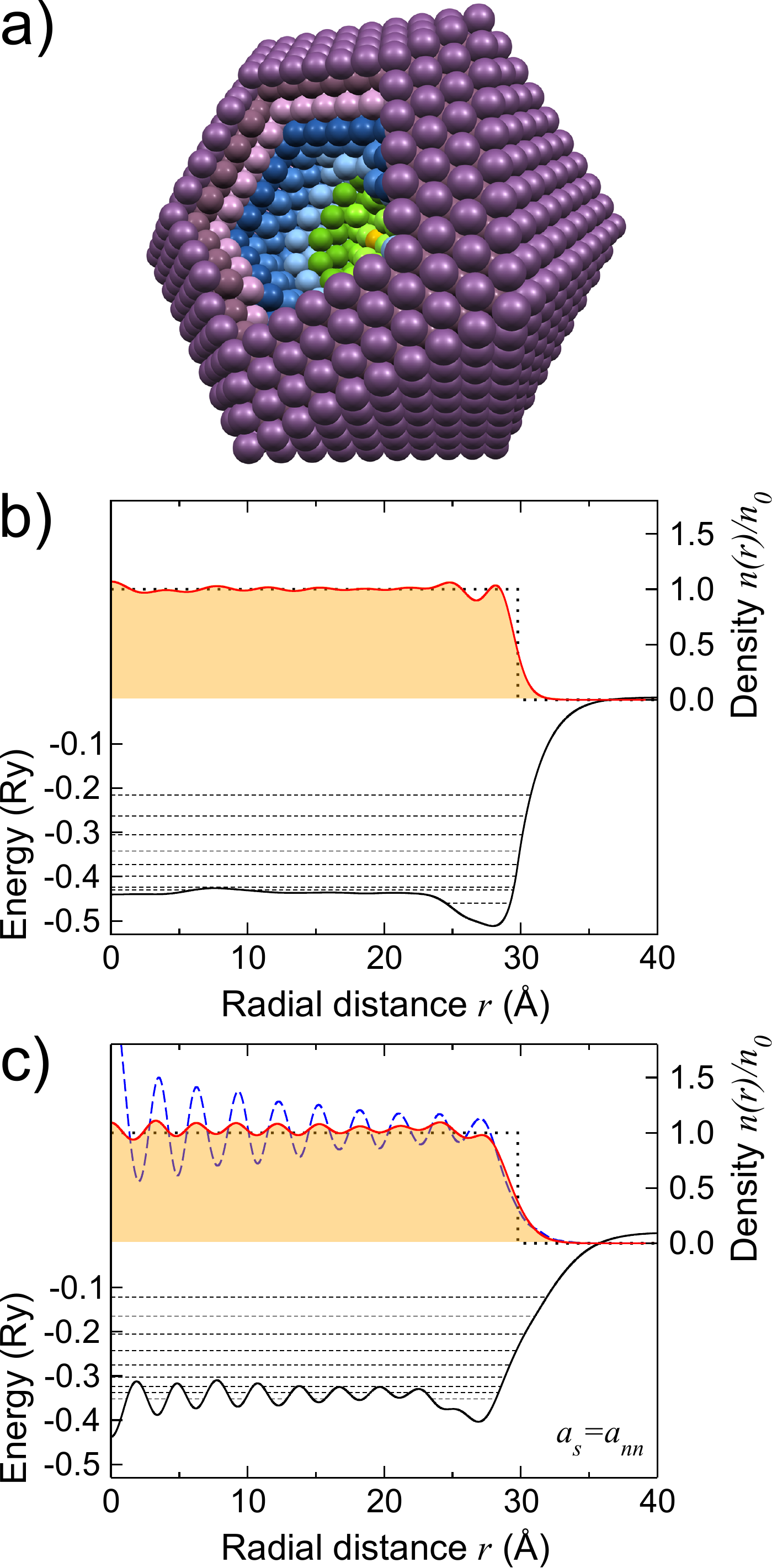}
\caption{(Color online) a) Geometrical structure of 10-shell icosahedral Na$_{2869}^-$
  cluster. The next two panels show the electronic structure (only $\ell=0$ states are
  shown), the Kohn-Sham potential (black solid line), the electron density (red solid
  line) from the self-consistent local density approximation (LDA) calculations for the
  jellium model, the dotted line marks the ideal jellium background (b) and for the
  spherically averaged realistic ionic density (dashed blue line) (c). The intershell
  spacing is fixed at the bulk nearest neighbor distance.
\label{fig:sys}}
\end{figure}
In the present contribution we continue to study the optical properties of the magic
Na$_{2869}^-$ cluster (Fig.~1a). This system simultaneously possesses completely closed
geometric and electronic shells. This makes it exceptionally stable and attractive for the
numerical calculations: its electronic structure can be obtained easily by using the
density functional approach. We do not pursue here a fully atomistic approach, but rather
solve the radial Kohn-Sham equation (using the renormalized \emph{Numerov method}) in the
presence of the \emph{spherically symmetric ionic density}. We either use the standard jellium
model which starts from the homogeneous ionic density (Fig.~1b) or the density is obtained
from the realistic geometric model by applying the gaussian broadening of $\sim 1$~{\AA}
width to ionic positions and performing a spherical averaging (Fig.~1c).  In accordance
with the proposed numerical scheme we present here a) the dipolar and the quadrupole
linear optical absorption coefficients $\alpha_{\ell 0}(\omega)$ and the induced densities
$\delta n^{(1)}_{\ell 0}(r;\omega)$ ($\ell=1,\,2$) (Fig.~2); b) results for the
second-harmonic source $\xi^{(2)}(\vec r)$; and c) the non-linear dipolar and quadrupole
optical responses at $2\omega$ frequency (Fig.~3).
\begin{figure}[hb!]
\includegraphics[width=\columnwidth]{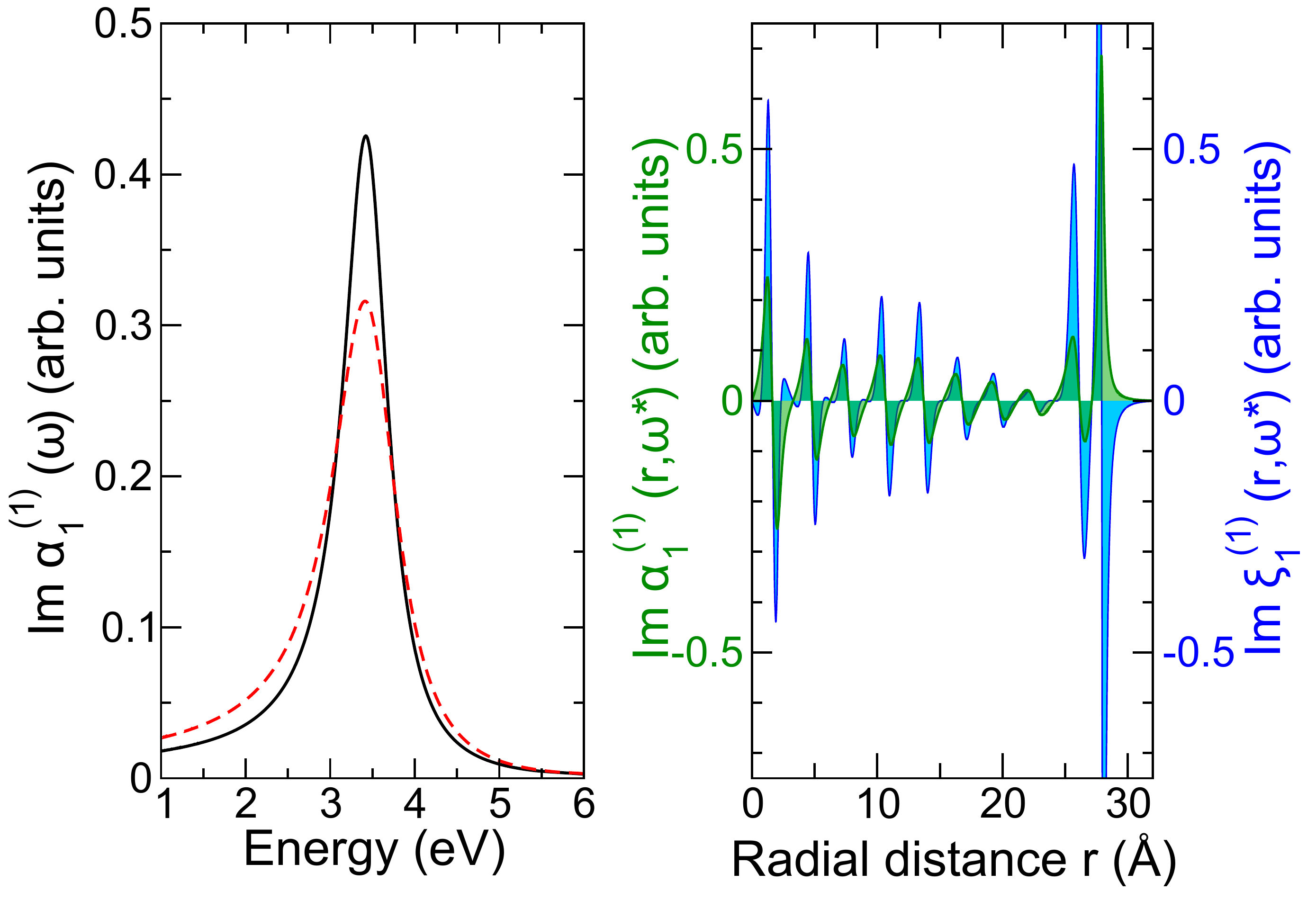}\\
\includegraphics[width=\columnwidth]{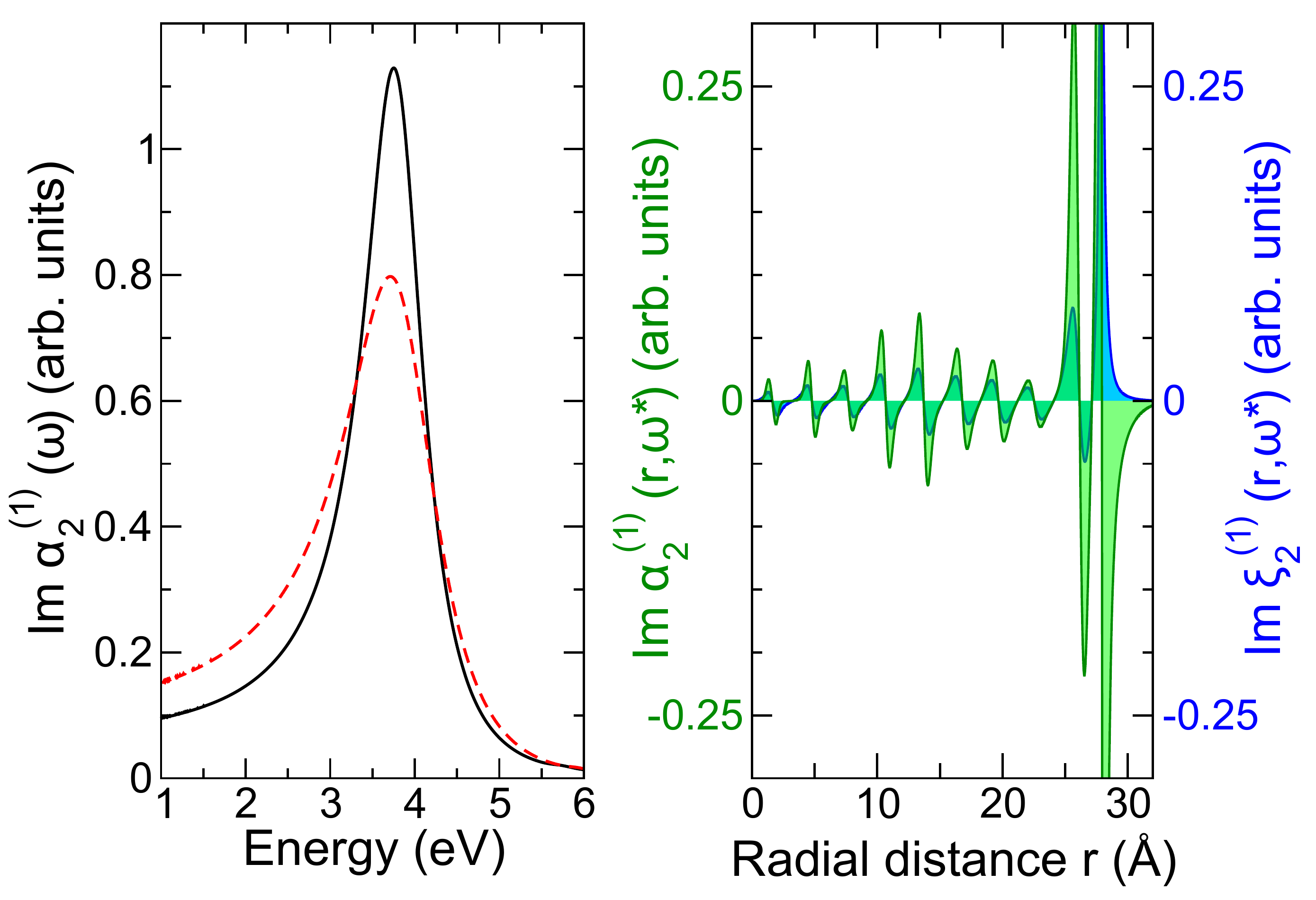}
\caption{(Color online) The linear $\ell=1$ (top row) and $\ell=2$ (bottom row) optical
  response of the icosahedral Na$_{2869}^-$ cluster based on the standard jellium model
  (black solid line) and on the model with the realistic ionic density (red dashed
  line). Right panels show the spatial dependence of the corresponding source terms
  $\xi^{(1)}_\ell(r,\omega^*)$ and the resulting induced densities $\delta
  n^{(1)}_\ell(r,\omega^*)$ for a particular frequency value $\omega^*=5.75$~eV.
\label{fig:zas1}}
\end{figure}
\begin{figure}[ht!]
\includegraphics[width=\columnwidth]{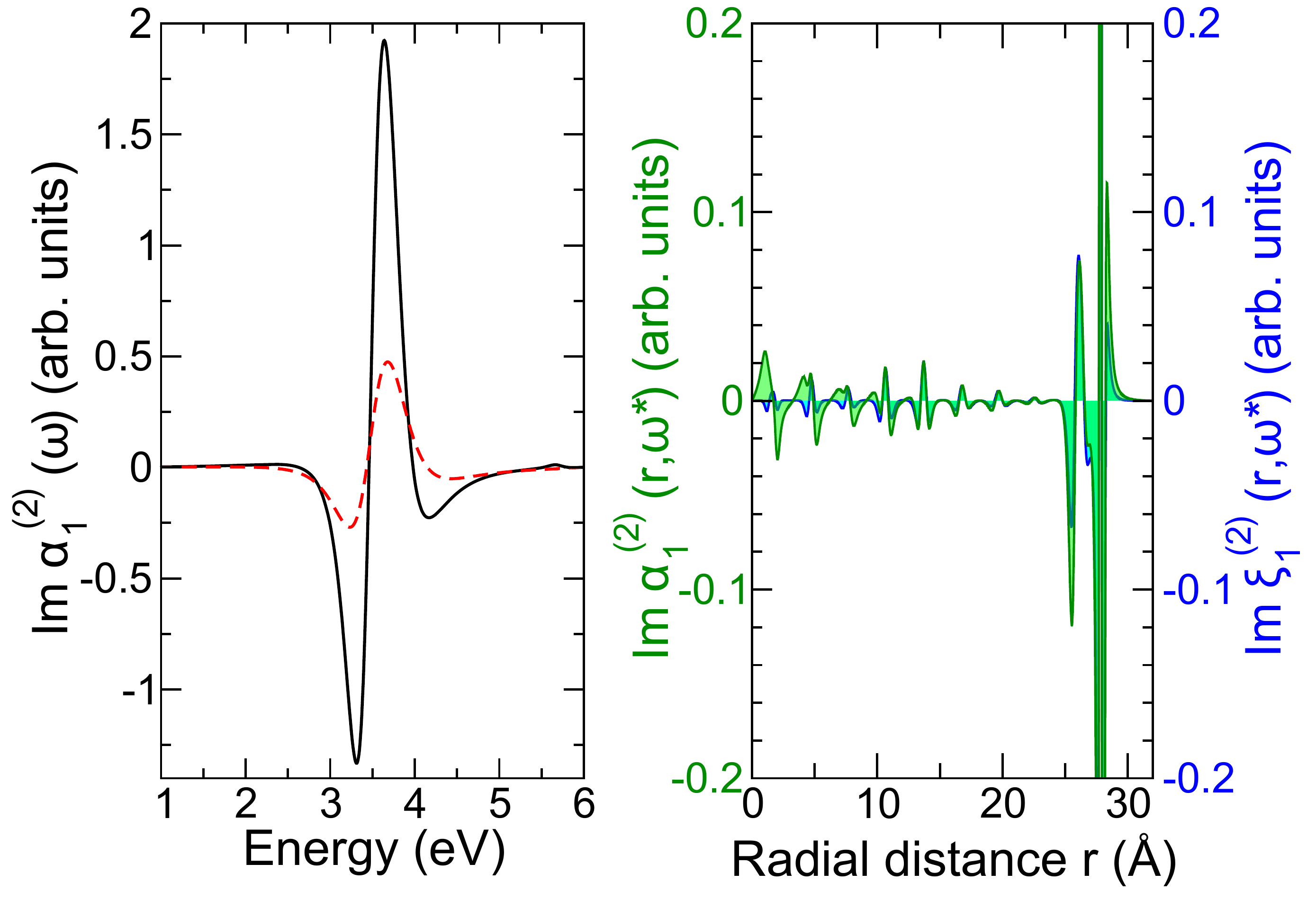}\\
\includegraphics[width=\columnwidth]{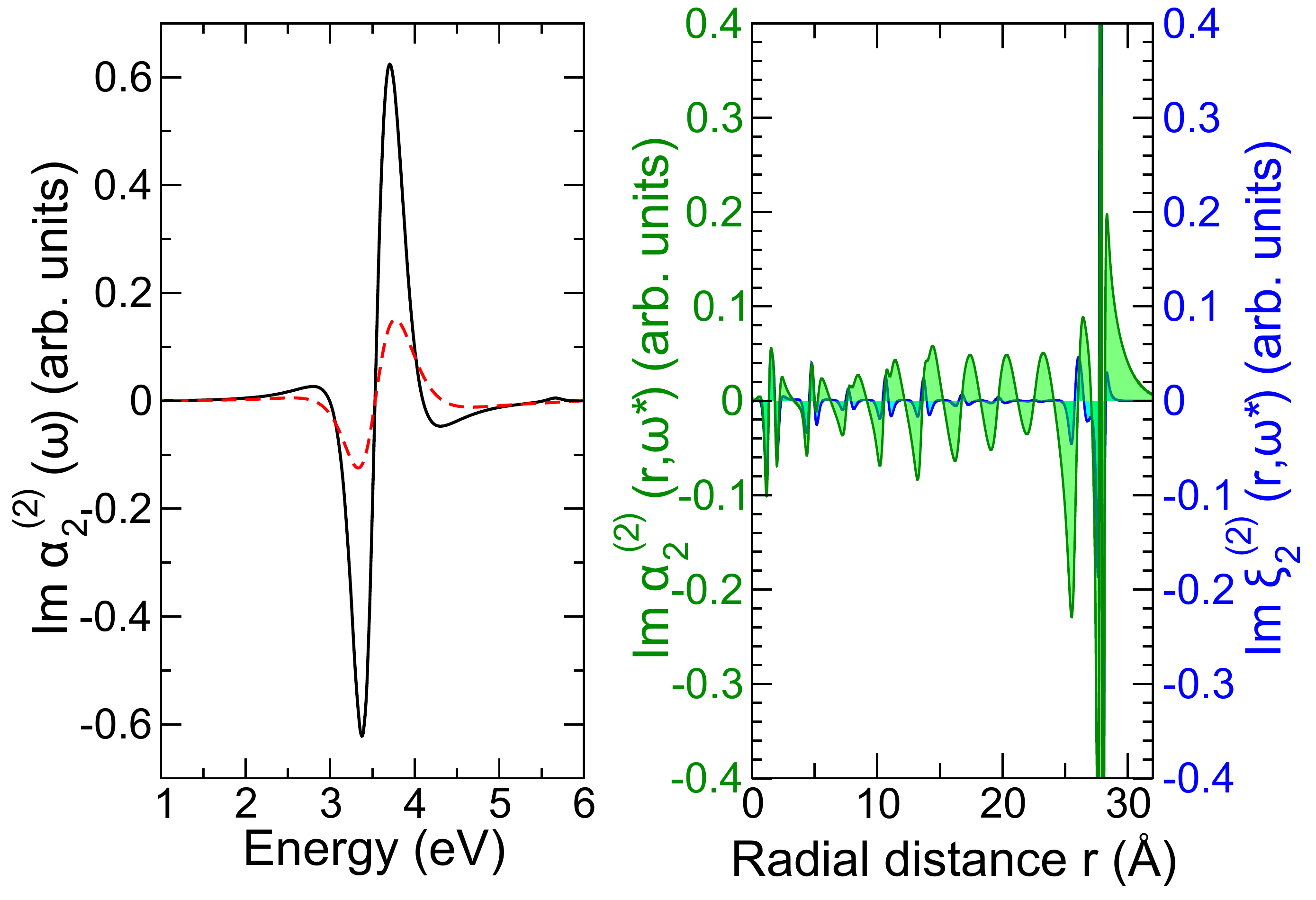}
\caption{(Color online) The second-order non-linear $\ell=1$ (top row) and $\ell=2$
  (bottom row) optical response of the icosahedral Na$_{2869}^-$ cluster for the two
  models at Fig.~2. Right panels show the spatial dependence of the corresponding source
  terms $\xi^{(2)}_\ell(r,\omega^*)$ and the resulting induced densities $\delta
  n^{(2)}_\ell(r,\omega^*)$ for a particular frequency value $\omega^*=5.75$~eV.
\label{fig:zas2}}
\end{figure}

While the linear optical response of metallic clusters is well understood: typically the
optical absorption profile only slightly deviates from that of the idealistic system: a
sphere filled with an electron gas of constant density. There, the optical absorption
coefficient is peaked at the energies of the surface plasmon modes:
\[
\omega_\ell=\sqrt{\frac{\ell}{2\ell+1}} \omega_p.
\]
 The position of the maxima is only weakly dependent on the details of the electronic
structure: our calculations almost perfectly match corresponding idealistic values
$\omega_1=3.45$~eV and $\omega_2=3.78$~eV for the bulk Na density ($r_s=3.96$). The
spill-off of the electron density in realistic systems mostly leads to the broadening of
the surface plasmon resonances. To illustrate this fact we choose the off-resonance value
of the frequency ($\omega^*=5.75$~eV) and plot the source $\xi^{(2)}_\ell(r,\omega^*)$ and
the induced density $\delta n^{(2)}_\ell(r,\omega^*)$ (Fig.~2). While in the idealistic
situation the induced density is distributed over the surface of the sphere (where the
electron density abruptly changes) in the realistic case we observe numerous features
associated with slow electronic density variations within the cluster. They contribute to
the optical absorption in the off-resonance regime. However, the relative weight of these
oscillations decreases  when the frequency approaches the resonance. There, the fast
density variation at the surface dominates the spectrum.

The non-linear optical properties (Fig.~3) of even such simple systems are not fully
understood. It is interesting to observe that two different excitation mechanisms (the
quadrupole transition at $\omega$ or $2\omega$ frequency) lead to almost identical
frequency dependence. Unlike in the linear case, the efficiency of the frequency
conversion vanishes at the plasmon resonance and has two pronounced peaks at the energy
slightly below and above. We also do not observe a strong correlation between the source
and the induced density as in the off-resonant linear case (cf. blue and green
curves). However, the spatial variation of these quantities is not erratic (as the plots
might be suggesting). The complicated radial dependence is the result of the derivatives
of the induced density and the potential in the non-linear source terms. Thanks to the
linear-scaling algorithms we are using at each stage of the computation we are able to
eliminate any spurious contributions from the numerical differentiation while ensuring the
convergence of our results with respect to the number of discretization points and the
value of the broadening parameter $\eta$.
\section{Conclusions}
We developed a semi-classical theory of the second-harmonic generation in spherical
particles. Although it originates from the exact quantum-mechanical sum-over-states
expression and takes the local fields into account it is free from the small system size
limitation. Because an efficient method for the solution of the corresponding integral
equations exists the theory can be applied even to macroscopic systems provided the
electronic density can be found. For this purpose the jellium approximation for the ionic
density can be used as we illustrate by computing the SHG spectrum of the Na$_{2869}^-$
cluster. The scattering cross-section and the angular distribution of the intensity are
other important experimental observables. Calculation of these quantities will be
presented elsewhere together with the inclusion of higher multipole moments allowing,
thus, for the treatment of larger systems. The work along these lines is already in
progress.

We believe that our method is sufficiently versatile as to allow for further
extensions. In particular, it is straightforward to modify the method for systems with
axial symmetry, or even to consider the symmetry-free case. To treat larger systems one
must include higher multipole moments.  This poses a question of how to find the
second-order source term in this case. We believe that a direct manipulation with the {\sc
maple} computer algebra system rather than a formal solution in terms of $3j$ symbols is
feasible.

Regarding  the physical aspects of our approach: The fact that it is free of any
adjustable parameters is not necessarily beneficial. In fact, for metallic systems with
localized $d$-electrons peculiarities of the electronic structure might be reflected in
the optical properties. In this case the classical electrodynamics approach with the
experimentally measured dielectric function and susceptibility tensor might give more
accurate results. On the other hand, for systems with simpler electronic structure our
method is capable of taking into account quantum effects such as the spill-out of the
electronic density.

Finally, our work establishes the  equivalence between the semi-classical approximation and
the hydrodynamic approach of Apell~\cite{apell_non-local_1983}. The latter, however, is
just a classical theory if not corrected for local field effects. We also touched upon the
question of the representability of the response functions in terms of electronic densities
and show that in general a more complicated quantity such as the one-particle density
matrix must be introduced. However, for the second-harmonic generation these terms cancel
in the final expression.

\acknowledgments The work is supported by DFG-SFB762 and DFG-SFB/TR 88. W.~H. gratefully
acknowledges the hospitality of the "Nonequilibrium Many-Body Systems" group at
Martin-Luther-University Halle-Wittenberg during his sabbatical.
\appendix
\section{Semi-classical expression for the first hyperpolarizability \label{sec:A}}
We start with the expression for the second order non-linear response:
\begin{multline}
\chi_2^{(0)}(\vec r;\vec r',\vec r'';\omega)=
\sum_{i,j,k}\frac{
\psi_k^*(\vec r)\psi_i(\vec r)
\psi_i^*(\vec r')\psi_j(\vec r')
\psi_j^*(\vec r'')\psi_k(\vec r'')
}{2\omega+E_i-E_k+i\eta}\\
\times\left(
\frac{f_k-f_j}{\omega+E_j-E_k+i\eta}-\frac{f_j-f_i}{\omega+E_i-E_j+i\eta}
\right)
\label{eq:chi2_gen}
\end{multline}
where $f$ is the Fermi function and $i$, $j$, $k$ refer to collections of quantum numbers
that uniquely characterize electronic states of the system. In what follows we use the
the following notations $E_i-E_k=E_{ik}$, $f_i-f_k=f_{ik}$, etc. The infinitesimally small
positive number $\eta$ shifts the poles from the real axis and assures, thus, the
causality of the response function. In what follows we will assume that it can be incorporated
in the $\omega$ variable.

Let us consider a generic type of integrals:
\begin{equation}
\Phi(\vec r)=\int\!d(\vec r'\vec r'')\chi_2^{(0)}(\vec r;\vec r',\vec r'';\omega)\,
\phi(\vec r')\,\phi(\vec r'')
\label{eq:phi}
\end{equation}
and introduce our basic approximation.  The semi-classical approximation (SCA) implies a high frequency condition
$|E_i-E_j|\ll \omega$.  Thus, the above expression can be expanded as a power series of
$\omega$.  The term proportional to $1/\omega^2$ in the expansion of $\Phi(\vec r)$
vanishes in view of the completeness of the electron wave-functions:
\begin{equation}
\sum_i\psi^*_i(\vec r_1)\psi_i(\vec r_2)=\delta(\vec r_1-\vec r_2).\label{eq:completeness}
\end{equation}
In what follows we will assume all wave-functions to be real. This can always be done
without the loss of generality for time-invariant systems.  Odd power terms vanish in view
of the symmetry of the expression with respect to permutation $i\rightleftarrows k$ and
$\vec r' \rightleftarrows\vec r''$.

The $1/\omega^4$ term has the following form:
\begin{multline}
\Phi^{(4)}(\vec r)=\int\!d(\vec r'\vec r'')\sum_{i,j,k}X_{i,j,k}(\vec r,\vec r',\vec r'')\,
\phi(\vec r')\,\phi(\vec r'')\\
\times\left[
f_{kj}E_{kj}^2-\frac12 E_{ik}f_{kj}E_{kj}+\frac14 E_{ik}^2f_{kj}
\right].
\end{multline}
The first term vanishes after using the property (\ref{eq:completeness}), the integration
over $\vec r'$, and exploiting the permutation symmetry $j\rightleftarrows k$ and $\vec r
\rightleftarrows\vec r''$ of the expression under the integral. The term proportional to
$E^2_{ik}f_j$ can be re-written as $2E_{ik}E_{jk}f_j$ and combined with the second
terms. Finally our expression can be written as:
\begin{equation}
\Phi^{(4)}(\vec r)=\frac14 \Phi^{(4a)}(\vec r)-\Phi^{(4b)}(\vec r)
+\frac12\Phi^{(4c)}(\vec r)
\label{eq:phi4_3terms}
\end{equation}
with following notations:
\begin{eqnarray*}
\Phi^{(4a)}(\vec r)&=&\int\!d(\vec r'\vec r'')\sum_{i,j,k}X_{i,j,k}(\vec r,\vec r',\vec r'')
\Big[E_{ik}^2f_{k}\Big]\phi(\vec r')\,\phi(\vec r''),\\
\Phi^{(4b)}(\vec r)&=&\int\!d(\vec r'\vec r'')\sum_{i,j,k}X_{i,j,k}(\vec r,\vec r',\vec r'')
\Big[E_{ik}E_{jk}f_{j}\Big]\phi(\vec r')\,\phi(\vec r''),\\
\Phi^{(4c)}(\vec r)&=&\int\!d(\vec r'\vec r'')\sum_{i,j,k}X_{i,j,k}(\vec r,\vec r',\vec r'')
\Big[E_{ik}E_{jk}f_{k}\Big]\phi(\vec r')\,\phi(\vec r'').
\end{eqnarray*}
In what follows we will make use of
\begin{equation}
E_{ij}\psi_i(\vec r)\psi_j(\vec r)
=-\frac12\vec\nabla \cdot \gvec{\xi}_{ij}(\vec r),
\label{eq:xi}
\end{equation}
where $\gvec \xi_{ij}(\vec r)=\psi_j(\vec{r})\vec\nabla \psi_i(\vec{r})
-\psi_i(\vec{r})\vec\nabla\psi_j(\vec{r})$. This follows from the Schr{\"o}dinger equation.

Let us consider first the $\Phi^{(4a)}(\vec r)$ term. After the summation over $j$ and
integration over $\vec r''$ we arrive at:
\[
\Phi^{(4a)}(\vec r)=\frac14\int\!d\vec r' \phi^2(\vec r') \sum_{i,k}f_{k}
[\vec\nabla_{r}\cdot \gvec \xi_{ik}(\vec r)]
[\vec\nabla_{r'}\cdot \gvec \xi_{ik}(\vec r')].
\]
We further pull $\vec\nabla_{r}$ out of the integration and use the Gauss theorem to apply
$\vec\nabla_{r'}$ to the $\phi^2(\vec r')$:
\begin{equation}
\Phi^{(4a)}=-\frac{\vec\nabla_{r}}{4}\cdot \int\!d\vec r'\sum_{i,k}f_{k} \,\gvec \xi_{ik}(\vec r)
\,[\gvec \xi_{ik}(\vec r')\cdot \vec\nabla_{r'}\phi^2(\vec r')].
\label{eq:phi4a_def}
\end{equation}
Now the summation over $i$ and $k$ can be performed by introducing the one-particle
density matrix:
\begin{equation}
\sum_i f_i\psi_i(\vec r_1)\psi_i(\vec r_2)=\gamma(\vec r_1,\vec r_2),\label{eq:1n}
\end{equation}
with the diagonal elements given by the density:
\begin{equation}
\sum_i f_i\psi_i(\vec r)\psi_i(\vec r)=n(\vec r),\label{eq:n}
\end{equation}
Quite general we can represent the density matrix in the form:
\[
\gamma(\vec r_1,\vec r_2)=f(\vec r_1)\,f(\vec r_2)g(|\vec r_1-\vec r_2|).
\]
From the definition \eqref{eq:1n} a useful integration formula can be derived:
\begin{multline}
\int\!d\vec r' \delta(\vec r-\vec r') F(\vec r') \vec\nabla_{r} \gamma(\vec r,\vec r')\\
=\int\!d\vec r' \delta(\vec r-\vec r') F(\vec r') \vec\nabla_{r'} \gamma(\vec r,\vec r')
=\frac12 F(\vec r) \vec\nabla_{r} n(\vec r).
\end{multline}
Let us now return to \eqref{eq:phi4a_def} and perform the sum over the states using
\cref{eq:1n,eq:n}.
\begin{multline}
\Phi^{(4a)}=-\frac{1}{4}\sum_{ab}\nabla_{r}^a\int\!d\vec r'\nabla_{r'}\phi^2(\vec r')\\
\times\Big[\gamma(\vec r,\vec r')\nabla_{r}^a\nabla_{r'}^b\delta(\vec r,\vec r')
-\nabla_{r}^a\gamma(\vec r,\vec r')\nabla_{r'}^b\delta(\vec r,\vec r')\\
-\nabla_{r}^a\delta(\vec r,\vec r')\nabla_{r'}^b\gamma(\vec r,\vec r')
+\delta(\vec r,\vec r')\nabla_{r}^a\nabla_{r'}^b\gamma(\vec r,\vec r')
\Big].
\end{multline}
In order to evaluate the integral we must remove all differential operators from the
$\delta$-function.
\begin{multline}
\Phi^{(4a)}=-\frac{1}{4}\sum_{ab}\nabla_{r}^a\int\!d\vec r'\nabla_{r'}^b\phi^2(\vec r')\\
\times\Big[\nabla_{r'}^b\nabla_{r}^a[\delta(\vec r,\vec r')\gamma(\vec r,\vec r')]
-2\nabla_{r}^a[\delta(\vec r,\vec r')\nabla_{r'}^b\gamma(\vec r,\vec r')]\\
-2\nabla_{r'}^b[\delta(\vec r,\vec r')\nabla_{r}^a\gamma(\vec r,\vec r')]
+4\delta(\vec r,\vec r')\nabla_{r}^a\nabla_{r'}^b\gamma(\vec r,\vec r')
\Big].
\end{multline}
It can be further simplified by using Green's first identity for the first and third
terms:
\begin{multline}
\Phi^{(4a)}=\frac{\Delta}{4}\int\!d\vec r'
\delta(\vec r,\vec r')\Big[\Delta\phi^2(\vec r')\gamma(\vec r,\vec r')\\
+2\nabla_{r'}^b\phi^2(\vec r')\nabla_{r'}^b\gamma(\vec r,\vec r')\Big]\\
-\frac12\sum_a\nabla_{r}^a\int\!d\vec r'\delta(\vec r,\vec r')
\Big[\Delta[\phi^2(\vec r')]\nabla_{r}^a\gamma(\vec r,\vec r')\\
+2\sum_b\nabla_{r}^a\nabla_{r'}^b\gamma(\vec r,\vec r')\nabla_{r'}^b\phi^2(\vec r')\Big].
\end{multline}
Let us introduce the tensor $\mathbb{T}^{ab}(\vec r)=\lim_{\vec r'\rightarrow\vec r}
\nabla_{r}^a\nabla_{r'}^b\gamma(\vec r,\vec r')$ and integrate:
\begin{multline}
\Phi^{(4a)}=\frac14\Delta\Big[n\,\Delta\phi^2+(\vec\nabla\phi^2
  \cdot\vec\nabla n)\Big]\\
-\frac14\vec\nabla\cdot\Big[\vec\nabla n\,\Delta\phi^2\Big]
-\sum_{ab}\nabla^a \Big[\mathbb{T}^{ab} \nabla^b\phi^2\Big].
\label{eq:phi4a}
\end{multline}

Using \eqref{eq:xi} we re-write $\Phi^{(4b)}$ in the form without explicit energies:
\begin{multline}
\Phi^{(4b)}(\vec r)=\frac{\vec \nabla_r}{4}\cdot\int\!d(\vec r'\vec r'')\,\phi(\vec r')\phi(\vec r'')
\sum_{i,j,k}f_{j} \gvec \xi_{ik}(r) \big[\vec \nabla_{r''}\cdot \gvec \xi_{jk}\big]\\
\times\psi_i(\vec r')\psi_j(\vec r').
\end{multline}
We perform the summations over the states and use the Gauss theorem to transfer $\vec
\nabla_{r''}$ on $\phi(\vec r'')$ to obtain
\begin{multline}
\Phi^{(4b)}(\vec r)=-\frac14\sum_{ab}\nabla_r^a\int\!d(\vec r'\vec r'')
\phi(\vec r')\,\nabla_{r''}^b\phi(\vec r'')\\
\times\Big[
\delta(\vec r,\vec r'')\nabla_r^a\delta(\vec r,\vec r')\,
\nabla_{r''}^b\gamma(\vec  r'',\vec r')\\
-\nabla_{r''}^b\delta(\vec r,\vec r'')\,\nabla_r^a\delta(\vec r,\vec r')\,
\gamma(\vec  r'',\vec r')\\
-\nabla_r^a\delta(\vec r,\vec r'')\,\delta(\vec r,\vec r')\,
\nabla_{r''}^b\gamma(\vec  r'',\vec r')\\
+\nabla_{r''}^b\nabla_r^a\Big(\delta(\vec r,\vec r'')\Big)\,\delta(\vec r,\vec r')\,
\gamma(\vec  r'',\vec r')
\Big]\label{eq:4b}
\end{multline}
The integration yields:
\begin{eqnarray*}
\Phi^{(4b_1)}(\vec r)&=&-\frac14\sum_{ab}\nabla_r^a\Big[\nabla_{r}^b\phi(\vec r)\nabla_r^a\int\!d\vec r'
\phi(\vec r')\delta(\vec r,\vec r')\,\nabla_{r}^b\gamma(\vec  r,\vec r')\\
&&+\frac14\sum_{ab}\nabla_r^a\Big[\nabla_{r}^b\phi(\vec r)\!\int\!d\vec r'
\phi(\vec  r')\,\delta(\vec r,\vec r')\,\nabla_r^a \nabla_{r}^b\gamma(\vec  r,\vec r')\Big]\\
&=&-\frac14\sum_{a}\nabla^a\Big[
\frac12(\nabla n\cdot \nabla \phi )\nabla^a\phi +\phi\sum_b\mathbb{T}^{ab}\nabla^b\phi
\Big],\\
\Phi^{(4b_3)}(\vec r)&=&\frac18\sum_{a}\nabla^a\Big[\phi\nabla^a(\nabla\phi\cdot\nabla n)
-2\phi\sum_{b}\mathbb{T}^{ab}\nabla^b\phi\Big],\\
\Phi^{(4b_4)}(\vec r)&=&-\frac14\sum_{ab}\nabla_r^a\Big[\phi(\vec r)\!\!\int\!d\vec r''
\nabla_{r''}^b\phi(\vec r'')\,\gamma(\vec r'',\vec r)\,\nabla_{r''}^b\nabla_r^a\delta(\vec r,\vec r'')\Big]\\
&=&-\frac14\sum_{ab}\nabla_r^a\Big[\phi(\vec r)\nabla_r^a\!\!\int\!d\vec r''\nabla_{r''}^b\phi(\vec r'')\,
\gamma(\vec r'',\vec r)\,\nabla_{r''}^b\delta(\vec r,\vec r'')\Big]\\
&+&\frac14\sum_{ab}\nabla_r^a\Big[\phi(\vec r)\!\int\!d\vec r''\nabla_{r''}^b\phi(\vec r'')\,
\nabla_r^a\gamma(\vec r'',\vec r)\,\nabla_{r''}^b\delta(\vec r,\vec r'')\Big].
\end{eqnarray*}
The $\Phi^{(4b_4)}(\vec r)$ term requires special attention:
\begin{eqnarray*}
\lefteqn{\Phi^{(4b_4)}(\vec r)}\\
&=&-\frac14\sum_{ab}\nabla_r^a\Big[\phi(\vec r)\nabla_r^a\!\int\!d\vec r''\nabla_{r''}^b\phi(\vec r'')\,
\nabla_{r''}^b\big[\delta(\vec r,\vec r'')\gamma(\vec r'',\vec r)\big]\Big]\\
&+&\frac14\sum_{ab}\nabla_r^a\Big[\phi(\vec r)\nabla_r^a\!\int\!d\vec r''\delta(\vec r,\vec r'')\,
\nabla_{r''}^b\phi(\vec r'')\,\nabla_{r''}^b\gamma(\vec r'',\vec r)\Big]\\
&+&\frac14\sum_{ab}\nabla_r^a\Big[\phi(\vec r)\!\int\!d\vec r''\nabla_{r''}^b\phi(\vec r'')\,
\nabla_{r''}^b\big[\delta(\vec r,\vec r'')\nabla_r^a\gamma(\vec r'',\vec r)\big]\Big]\\
&-&\frac14\sum_{ab}\nabla_r^a\Big[\phi(\vec r)\!\int\!d\vec r''\delta(\vec r,\vec r'')\,
\nabla_{r''}^b \phi(\vec r'')\,\nabla_{r''}^b\nabla_r^a\gamma(\vec r'',\vec r)\Big]\\
&=&\frac18\sum_{a}\nabla^a\Big[\phi\nabla^a(\nabla\phi\cdot \nabla n)
-2\sum_b\phi\mathbb{T}^{ab}\nabla^b\phi\Big]\\
&+&\frac18\sum_{a}\nabla^a\Big[2\phi\nabla^a\big(n\Delta\phi\big)-\phi\Delta \phi \nabla^a n\Big].
\end{eqnarray*}
\begin{eqnarray*}
\lefteqn{\Phi^{(4b_2)}(\vec r)}\\
&=&\frac14\Delta\sum_{b}\int\!d(\vec r'\vec r'')\,\phi(\vec r')\,
\nabla_{r''}^b\phi(\vec r'')\,\nabla_{r''}^b\delta(\vec r,\vec r'')\\
&\times&\delta(\vec r,\vec r')\,\gamma(\vec r'',\vec r')+\Phi^{(4b_4)}(\vec r)\\
&=&\frac14\Delta\left[\phi\!\!\int\!d\vec r''\sum_{b}\nabla_{r''}^b\phi(\vec r'')\,
\nabla_{r''}^b\Big[\delta(\vec r,\vec r'')\gamma(\vec  r'',\vec r)\Big]\right]\\
&-&\frac14\Delta\left[\phi\!\!\int\!d\vec r''\sum_{b}\nabla_{r''}^b\phi(\vec r'')\,
\nabla_{r''}^b\gamma(\vec  r'',\vec r)\,\delta(\vec r,\vec r'')\right]+\Phi^{(4b_4)}(\vec r)\\
&=&-\frac14\Delta\Big[\phi n\Delta\phi+\frac12\phi(\nabla n\cdot\nabla\phi)\Big]
+\Phi^{(4b_4)}(\vec r).
\end{eqnarray*}
Combining together we obtain:
\begin{multline}
\Phi^{(4b)}(\vec r)
=-\frac14\Delta\Big[\phi\,(\nabla\phi\cdot\nabla n)\Big]
+\frac12\vec\nabla\cdot\Big[\phi\vec\nabla\big(\vec \nabla n\cdot\vec\nabla\phi\big)\Big]\\
-\frac14\Delta\Big[\phi n\Delta\phi\Big]
+\frac12\vec\nabla\cdot\Big[\phi\vec\nabla(n\Delta\phi)\Big]\\
-\sum_a\nabla^a\Big[\phi \mathbb{T}^{ab}\nabla^b \phi\Big]
-\frac14\vec\nabla\cdot\Big[\phi\Delta\phi\vec \nabla n\Big].
\label{eq:phi4b}
\end{multline}
Equation for $\Phi^{(4c)}$ is obtained from \eqref{eq:4b} with replacement $\delta(\vec
r,\vec r'')\rightarrow\gamma(\vec r,\vec r'')$ and $\gamma(\vec r'',\vec
r')\rightarrow\delta(\vec r'',\vec r')$:
\begin{multline}
\Phi^{(4c)}(\vec r)=-\frac14\sum_{ab}\nabla_r^a\int\!d(\vec r'\vec r'')
\phi(\vec r')\,\nabla_{r''}^b\phi(\vec r'')\\
\times\Big[
\gamma(\vec r,\vec r'')\,\nabla_r^a\delta(\vec r,\vec r')\,\nabla_{r''}^b\delta(\vec  r'',\vec r')\\
-\nabla_{r''}^b\gamma(\vec r,\vec r'')\,\nabla_r^a\delta(\vec r,\vec r')\,\delta(\vec  r'',\vec r')\\
-\nabla_r^a\gamma(\vec r,\vec r'')\,\delta(\vec r,\vec r')\,\nabla_{r''}^b\delta(\vec  r'',\vec r')\\
+\nabla_{r''}^b\nabla_r^a\Big(\gamma(\vec r,\vec r'')\Big)\,\delta(\vec r,\vec r')\,
\delta(\vec  r'',\vec r')\Big].
\end{multline}
Integrations yield:
\begin{eqnarray*}
\Phi^{(4c_4)}(\vec r)&=&-\frac14\sum_{ab}\nabla^a\Big[\phi\mathbb{T}^{ab}\nabla^b\phi\Big],\\
\Phi^{(4c_3)}(\vec r)&=&-\frac14\sum_{ab}\nabla_r^a\phi\int\!d\vec r'' \nabla_{r''}^b\phi(\vec r'')\,
\nabla_r^a\gamma(\vec r,\vec r'')\,\nabla_{r''}^b\delta(\vec  r'',\vec r)\\
&=&+\frac18\sum_{a}\nabla^a\Big[\phi \Delta\phi \nabla^a n+2\phi\sum_b\mathbb{T}^{ab}\nabla^b\phi\Big],\\
\Phi^{(4c_2)}(\vec r)&=&\frac14\sum_{ab}\nabla_r^a\int\!d\vec r'\,\phi(\vec r')\,
\nabla_{r'}^b\phi(\vec r')\,\nabla_{r'}^b\gamma(\vec r,\vec r')\,\nabla_r^a\delta(\vec r,\vec r')\\
&=&\frac{1}{8}\Delta\Big[\phi(\nabla\phi\cdot\nabla n)\Big]
-\frac14\sum_{ab}\nabla^a\Big[\phi\mathbb{T}^{ab}\nabla^b\phi\Big].
\end{eqnarray*}
It is little bit more difficult to get the first term:
\begin{eqnarray*}
\Phi^{(4c_1)}(\vec r)&=&-\frac14\Delta\Big[\phi\int\!d\vec r''\,\gamma(\vec r,\vec r'')
\sum_b\nabla_{r''}^b\phi(\vec r'')\,\nabla_{r''}^b\delta(\vec  r'',\vec r)\Big]\\
&+&\frac14\sum_{ab}\nabla_r^a\phi\int\!d\vec r''\,
\nabla_{r''}^b\phi(\vec r'')\,\nabla_r^a\gamma(\vec r,\vec r'')\,
\nabla_{r''}^b\delta(\vec  r'',\vec r)\\
&=&\frac14\Delta\Big[\phi n\Delta \phi\Big]+\frac18\Delta\Big[\phi(\nabla\phi\cdot\nabla n)\Big]\\
&-&\frac18\sum_a\nabla^a\Big[\phi \Delta\phi\nabla^a n\Big]
-\frac14 \sum_{ab}\nabla^a\Big[\phi \mathbb{T}^{ab}\nabla^b \phi\Big].
\end{eqnarray*}
Combining together we obtain:
\begin{multline}
\Phi^{(4c)}(\vec r)
=\frac14\Delta\Big[\phi\,(\nabla\phi\cdot\nabla n)\Big]
+\frac14\Delta\Big[\phi n\Delta\phi\Big]\\
-\sum_a\nabla^a\Big[\phi \mathbb{T}^{ab}\nabla^b \phi\Big]
-\frac14\vec\nabla\cdot\Big[\phi\Delta\phi\vec \nabla n\Big].
\label{eq:phi4c}
\end{multline}
Finally substituting \cref{eq:phi4a,eq:phi4b,eq:phi4c} in \eqref{eq:phi4_3terms} we
obtain for the forth order term:
\begin{equation}
\Phi^{(4)}(\vec r)=\frac12\vec\nabla\cdot\Big[\vec\nabla\phi\, \Big(n\Delta \phi+(\nabla\phi\cdot\nabla n)\Big)
+\frac14 n\vec\nabla\big(\vec\nabla\phi\big)^2\Big].\label{eq:phi4_fin}
\end{equation}
$\Phi(\vec r)$ has the meaning of an induced density, i.e. it is a gradient of the
non-linear polarization. Expression for this quantity was derived by Wang, Cheng and
Bower~\cite{wang_second_1973} and Apell~\cite{apell_non-local_1983} from the classical
equation of motion for the electron in electromagnetic fields. Our expression is identical
to their result (cf. Eq.~(11) in \cite{wang_second_1973,apell_non-local_1983}) if we write
the external field in the form $\vec E=-\vec\nabla\phi$. Our expression is more general in
a sense that $\vec E$ also contains the induced fields. Note there are no more terms
containing $\gamma(\vec r,\vec r'')$ in the expression \eqref{eq:phi4_fin}.  The
high-frequency condition is not the only way to simplify Eq.~\eqref{eq:chi2_gen}. In
particular, Rudnick and Stern~\cite{rudnick_second-harmonic_1971} demonstrated another
possibility to arrive at the classical expression (Eq.~\eqref{eq:apell}) starting from the
quantum mechanical response function for the homogeneous electron gas and applying the
local approximation $qv_F\ll \omega$, where $v_F$ is the Fermi velocity and $q$ is the
inverse of the characteristic length of variation of the field.
\section{Non-linear source term $\xi^{(2b)}_{\ell 0}(r)$}
Here we present expressions for the second part of the second-harmonic source
term. Calculations are facilitated by the use of {\sc maple} computer algebra package.
\begin{multline}
\xi^{(2b)}_1=-\frac{1}{4\omega^4\sqrt{5\pi}}\Bigg[
\frac{6f_1f_2}{r^4}\Big(rn'-8n\Big)+\frac{24 n}{r^3}\Big(f_1f_2'+2f_2f_1'\Big)\\
+\frac{4\pi n}{r^2}\Big(7f_1n_2+15n_1f_2\Big)
-\frac{n'}{r^2}\Big(5f_1f_2'+9f_2f_1'\Big)\\
+\frac{4\pi}{r}\Big(r n'-4n\Big)\Big(n_1f_2'+n_2f_1'+\frac{f_1'f_2'}{\pi r}\Big)\\
+4\pi n\Big(f_1'n_2'+f_2'n_1'-8\pi n_1n_2\Big)
\Bigg],
\end{multline}
\begin{multline}
\xi^{(2b)}_2=-\frac{1}{4\omega^4\sqrt{5\pi}}\Bigg[
4\pi n \Big(f_1'n_1'-4\pi n_1^2\Big)+\frac{4\pi n}{r^3}\Big(3 r f_1n_1-4f_1'n_1\Big)\\
+\frac{8nf_1'f_1}{r^3}-\frac{4n}{r^4}\Big(f_1^2+r^2(f_1')^2\Big)\\
+n'\Big(
4\pi n_1f_1'+\frac{2(f_1')^2}{r}-\frac{f_1f_1'}{r^2}-\frac{f_1^2}{r^3}
\Big)
\Bigg].
\end{multline}
We use the same abbreviated notations as in Sec. III and additionally introduce
$\xi^{(2b)}_\ell\equiv\xi^{(2b)}_{\ell 0}$, $n\equiv n(r)$.
\section{Fast numerical solution of the integral equation}
It is convenient to recast \cref{eq:int_nsph,eq:int_asph} in the following form:
\begin{multline}
\alpha_\ell(r;\omega)=\xi_\ell(r;\omega)-\alpha^{(0)}_\ell(r;\omega)\\
\times\int_0^\infty d r'
\left(\frac{1}{r'}\right)^{\ell-1}\!G^\text{sph}_\ell(r,r')\,
\alpha_\ell(r';\omega),
\end{multline}
where for the linear response we have
$\xi_\ell(r;\omega)=\ell\alpha^{(0)}_\ell(r;\omega)$. Let us now omit the $\omega$
argument for simpification and use the definition \eqref{eq:G_sph}:
\[
\alpha_\ell(r)=\xi_\ell(r)+\alpha^{(0)}_\ell(r)
\Big[\ell\,a-\frac{\ell+1}{r^{2\ell+1}}w_1(r)-\ell w_2(r)\Big],
\]
with
\begin{eqnarray*}
w_1(r)&=&\int_0^r \Big(r'\Big)^{\ell+2} \alpha_\ell(r')\,dr',\\
w_2(r)&=&\int_0^r \left(\frac1{r'}\right)^{\ell-1}\!\!\!\alpha_\ell(r')\, dr',\quad
\text{and}\quad a=w_2(\infty).
\end{eqnarray*}
Functions $w_{1,2}(r)$ satisfy following differential equations:
\begin{eqnarray*}
w'_1(r)&=&r^{\ell+2}\Bigg\{\xi_\ell(r)+\alpha^{(0)}_\ell(r)\,
\Big[\ell\,a-\frac{\ell+1}{r^{2\ell+1}}w_1(r)-\ell w_2(r)\Big]\Bigg\},\\
w'_2(r)&=&\frac1{r^{\ell-1}}\Bigg\{\xi_\ell(r)+\alpha^{(0)}_\ell(r)\,
\Big[\ell\,a-\frac{\ell+1}{r^{2\ell+1}}w_1(r)-\ell w_2(r)\Big]\Bigg\}.
\end{eqnarray*}
They can be solved at the $\mathcal{O}(N)$ cost where $N$ is the number of mesh points to
represent the density.

\begin{thebibliography}{39}%
\makeatletter
\providecommand \@ifxundefined [1]{%
 \@ifx{#1\undefined}
}%
\providecommand \@ifnum [1]{%
 \ifnum #1\expandafter \@firstoftwo
 \else \expandafter \@secondoftwo
 \fi
}%
\providecommand \@ifx [1]{%
 \ifx #1\expandafter \@firstoftwo
 \else \expandafter \@secondoftwo
 \fi
}%
\providecommand \natexlab [1]{#1}%
\providecommand \enquote  [1]{``#1''}%
\providecommand \bibnamefont  [1]{#1}%
\providecommand \bibfnamefont [1]{#1}%
\providecommand \citenamefont [1]{#1}%
\providecommand \href@noop [0]{\@secondoftwo}%
\providecommand \href [0]{\begingroup \@sanitize@url \@href}%
\providecommand \@href[1]{\@@startlink{#1}\@@href}%
\providecommand \@@href[1]{\endgroup#1\@@endlink}%
\providecommand \@sanitize@url [0]{\catcode `\\12\catcode `\$12\catcode
  `\&12\catcode `\#12\catcode `\^12\catcode `\_12\catcode `\%12\relax}%
\providecommand \@@startlink[1]{}%
\providecommand \@@endlink[0]{}%
\providecommand \url  [0]{\begingroup\@sanitize@url \@url }%
\providecommand \@url [1]{\endgroup\@href {#1}{\urlprefix }}%
\providecommand \urlprefix  [0]{URL }%
\providecommand \Eprint [0]{\href }%
\@ifxundefined \urlstyle {%
  \providecommand \doi  [0]{\begingroup \@sanitize@url \@doi}%
  \providecommand \@doi [1]{\endgroup \@@startlink {\doibase
  #1}doi:\discretionary {}{}{}#1\@@endlink }%
}{%
  \providecommand \doi  [0]{doi:\discretionary{}{}{}\begingroup
  \urlstyle{rm}\Url }%
}%
\providecommand \doibase [0]{http://dx.doi.org/}%
\providecommand \Doi [0]{\begingroup \@sanitize@url \@Doi }%
\providecommand \@Doi  [1]{\endgroup\@@startlink{\doibase#1}\@@Doi}%
\providecommand \@@Doi [1]{#1\@@endlink}%
\providecommand \selectlanguage [0]{\@gobble}%
\providecommand \bibinfo  [0]{\@secondoftwo}%
\providecommand \bibfield  [0]{\@secondoftwo}%
\providecommand \translation [1]{[#1]}%
\providecommand \BibitemOpen [0]{}%
\providecommand \bibitemStop [0]{}%
\providecommand \bibitemNoStop [0]{.\EOS\space}%
\providecommand \EOS [0]{\spacefactor3000\relax}%
\providecommand \BibitemShut  [1]{\csname bibitem#1\endcsname}%
\bibitem [{\citenamefont {Mahan}(2000)}]{mahan_many-particle_2000}%
  \BibitemOpen
  \bibfield  {author} {\bibinfo {author} {\bibfnamefont {G.}~\bibnamefont
  {Mahan}},\ }\href@noop {} {\emph {\bibinfo {title} {Many-particle
  physics}}},\ \bibinfo {edition} {3rd}\ ed.\ (\bibinfo  {publisher} {Kluwer
  {Academic/Plenum} Publishers},\ \bibinfo {address} {New York},\ \bibinfo
  {year} {2000})\BibitemShut {NoStop}%
\bibitem [{\citenamefont {Halas}\ \emph {et~al.}(2011)\citenamefont {Halas},
  \citenamefont {Lal}, \citenamefont {Chang}, \citenamefont {Link},\ and\
  \citenamefont {Nordlander}}]{halas_plasmons_2011}%
  \BibitemOpen
  \bibfield  {author} {\bibinfo {author} {\bibfnamefont {N.~J.}\ \bibnamefont
  {Halas}}, \bibinfo {author} {\bibfnamefont {S.}~\bibnamefont {Lal}}, \bibinfo
  {author} {\bibfnamefont {W.}~\bibnamefont {Chang}}, \bibinfo {author}
  {\bibfnamefont {S.}~\bibnamefont {Link}}, \ and\ \bibinfo {author}
  {\bibfnamefont {P.}~\bibnamefont {Nordlander}},\ }\href@noop {} {\bibfield
  {journal} {\bibinfo  {journal} {Chem. Rev.},\ }\textbf {\bibinfo {volume}
  {111}},\ \bibinfo {pages} {3913} (\bibinfo {year} {2011})}\BibitemShut
  {NoStop}%
\bibitem [{\citenamefont {Roke}(2009)}]{roke_nonlinear_2009}%
  \BibitemOpen
  \bibfield  {author} {\bibinfo {author} {\bibfnamefont {S.}~\bibnamefont
  {Roke}},\ }\href@noop {} {\bibfield  {journal} {\bibinfo  {journal}
  {{ChemPhysChem}},\ }\textbf {\bibinfo {volume} {10}},\ \bibinfo {pages}
  {1380} (\bibinfo {year} {2009})}\BibitemShut {NoStop}%
\bibitem [{\citenamefont {Garc\'{i}a~de
  Abajo}(2010)}]{garcia_de_abajo_optical_2010}%
  \BibitemOpen
  \bibfield  {author} {\bibinfo {author} {\bibfnamefont {F.~J.}\ \bibnamefont
  {Garc\'{i}a~de Abajo}},\ }\href@noop {} {\bibfield  {journal} {\bibinfo
  {journal} {Rev. Mod. Phys.},\ }\textbf {\bibinfo {volume} {82}},\ \bibinfo
  {pages} {209} (\bibinfo {year} {2010})}\BibitemShut {NoStop}%
\bibitem [{\citenamefont {Liebsch}(1997)}]{liebsch_electronic_1997}%
  \BibitemOpen
  \bibfield  {author} {\bibinfo {author} {\bibfnamefont {A.}~\bibnamefont
  {Liebsch}},\ }\href@noop {} {\emph {\bibinfo {title} {Electronic excitations
  at metal surfaces}}},\ Physics of solids and liquids\ (\bibinfo  {publisher}
  {Plenum Press},\ \bibinfo {address} {New York},\ \bibinfo {year}
  {1997})\BibitemShut {NoStop}%
\bibitem [{\citenamefont {Andersen}\ and\ \citenamefont
  {H\"{u}bner}(2002)}]{andersen_substrate_2002}%
  \BibitemOpen
  \bibfield  {author} {\bibinfo {author} {\bibfnamefont {T.}~\bibnamefont
  {Andersen}}\ and\ \bibinfo {author} {\bibfnamefont {W.}~\bibnamefont
  {H\"{u}bner}},\ }\href@noop {} {\bibfield  {journal} {\bibinfo  {journal}
  {Phys. Rev. B},\ }\textbf {\bibinfo {volume} {65}},\ \bibinfo {pages}
  {174409} (\bibinfo {year} {2002})}\BibitemShut {NoStop}%
\bibitem [{\citenamefont {\"{O}stling}\ \emph {et~al.}(1993)\citenamefont
  {\"{O}stling}, \citenamefont {Stampfli},\ and\ \citenamefont
  {Bennemann}}]{oestling_theory_1993-1}%
  \BibitemOpen
  \bibfield  {author} {\bibinfo {author} {\bibfnamefont {D.}~\bibnamefont
  {\"{O}stling}}, \bibinfo {author} {\bibfnamefont {P.}~\bibnamefont
  {Stampfli}}, \ and\ \bibinfo {author} {\bibfnamefont {K.~H.}\ \bibnamefont
  {Bennemann}},\ }\href@noop {} {\bibfield  {journal} {\bibinfo  {journal} {Z.
  Phys. D: At. Mol. Clusters},\ }\textbf {\bibinfo {volume} {28}},\ \bibinfo
  {pages} {169} (\bibinfo {year} {1993})}\BibitemShut {NoStop}%
\bibitem [{\citenamefont {Dewitz}\ \emph {et~al.}(1996)\citenamefont {Dewitz},
  \citenamefont {H\"{u}bner},\ and\ \citenamefont
  {Bennemann}}]{dewitz_theory_1996}%
  \BibitemOpen
  \bibfield  {author} {\bibinfo {author} {\bibfnamefont {J.}~\bibnamefont
  {Dewitz}}, \bibinfo {author} {\bibfnamefont {W.}~\bibnamefont {H\"{u}bner}},
  \ and\ \bibinfo {author} {\bibfnamefont {K.}~\bibnamefont {Bennemann}},\
  }\href@noop {} {\bibfield  {journal} {\bibinfo  {journal} {Z. Phys. D: At.
  Mol. Clusters},\ }\textbf {\bibinfo {volume} {37}},\ \bibinfo {pages} {75}
  (\bibinfo {year} {1996})}\BibitemShut {NoStop}%
\bibitem [{\citenamefont {Dadap}\ \emph {et~al.}(1999)\citenamefont {Dadap},
  \citenamefont {Shan}, \citenamefont {Eisenthal},\ and\ \citenamefont
  {Heinz}}]{dadap_second-harmonic_1999}%
  \BibitemOpen
  \bibfield  {author} {\bibinfo {author} {\bibfnamefont {J.~I.}\ \bibnamefont
  {Dadap}}, \bibinfo {author} {\bibfnamefont {J.}~\bibnamefont {Shan}},
  \bibinfo {author} {\bibfnamefont {K.~B.}\ \bibnamefont {Eisenthal}}, \ and\
  \bibinfo {author} {\bibfnamefont {T.~F.}\ \bibnamefont {Heinz}},\ }\href@noop
  {} {\bibfield  {journal} {\bibinfo  {journal} {Phys. Rev. Lett.},\ }\textbf
  {\bibinfo {volume} {83}},\ \bibinfo {pages} {4045} (\bibinfo {year}
  {1999})}\BibitemShut {NoStop}%
\bibitem [{\citenamefont {Pavlyukh}\ and\ \citenamefont
  {H\"{u}bner}(2004)}]{pavlyukh_nonlinear_2004}%
  \BibitemOpen
  \bibfield  {author} {\bibinfo {author} {\bibfnamefont {Y.}~\bibnamefont
  {Pavlyukh}}\ and\ \bibinfo {author} {\bibfnamefont {W.}~\bibnamefont
  {H\"{u}bner}},\ }\href@noop {} {\bibfield  {journal} {\bibinfo  {journal}
  {Phys. Rev. B},\ }\textbf {\bibinfo {volume} {70}},\ \bibinfo {pages}
  {245434} (\bibinfo {year} {2004})}\BibitemShut {NoStop}%
\bibitem [{\citenamefont {de~Beer}\ and\ \citenamefont
  {Roke}(2009)}]{de_beer_nonlinear_2009}%
  \BibitemOpen
  \bibfield  {author} {\bibinfo {author} {\bibfnamefont {A.~G.~F.}\
  \bibnamefont {de~Beer}}\ and\ \bibinfo {author} {\bibfnamefont
  {S.}~\bibnamefont {Roke}},\ }\href@noop {} {\bibfield  {journal} {\bibinfo
  {journal} {Phys. Rev. B},\ }\textbf {\bibinfo {volume} {79}},\ \bibinfo
  {pages} {155420} (\bibinfo {year} {2009})}\BibitemShut {NoStop}%
\bibitem [{\citenamefont {Dadap}(2008)}]{dadap_optical_2008}%
  \BibitemOpen
  \bibfield  {author} {\bibinfo {author} {\bibfnamefont {J.~I.}\ \bibnamefont
  {Dadap}},\ }\href@noop {} {\bibfield  {journal} {\bibinfo  {journal} {Phys.
  Rev. B},\ }\textbf {\bibinfo {volume} {78}},\ \bibinfo {pages} {205322}
  (\bibinfo {year} {2008})}\BibitemShut {NoStop}%
\bibitem [{\citenamefont {de~Beer}\ \emph {et~al.}(2011)\citenamefont
  {de~Beer}, \citenamefont {Roke},\ and\ \citenamefont
  {Dadap}}]{de_beer_theory_2011}%
  \BibitemOpen
  \bibfield  {author} {\bibinfo {author} {\bibfnamefont {A.~G.~F.}\
  \bibnamefont {de~Beer}}, \bibinfo {author} {\bibfnamefont {S.}~\bibnamefont
  {Roke}}, \ and\ \bibinfo {author} {\bibfnamefont {J.~I.}\ \bibnamefont
  {Dadap}},\ }\href@noop {} {\bibfield  {journal} {\bibinfo  {journal} {J. Opt.
  Soc. Am. B},\ }\textbf {\bibinfo {volume} {28}},\ \bibinfo {pages} {1374}
  (\bibinfo {year} {2011})}\BibitemShut {NoStop}%
\bibitem [{\citenamefont {Prodan}\ and\ \citenamefont
  {Nordlander}(2003)}]{prodan_structural_2003}%
  \BibitemOpen
  \bibfield  {author} {\bibinfo {author} {\bibfnamefont {E.}~\bibnamefont
  {Prodan}}\ and\ \bibinfo {author} {\bibfnamefont {P.}~\bibnamefont
  {Nordlander}},\ }\href@noop {} {\bibfield  {journal} {\bibinfo  {journal}
  {Nano Lett.},\ }\textbf {\bibinfo {volume} {3}},\ \bibinfo {pages} {543}
  (\bibinfo {year} {2003})}\BibitemShut {NoStop}%
\bibitem [{\citenamefont {Orr}\ and\ \citenamefont
  {Ward}(1971)}]{orr_perturbation_1971}%
  \BibitemOpen
  \bibfield  {author} {\bibinfo {author} {\bibfnamefont {B.}~\bibnamefont
  {Orr}}\ and\ \bibinfo {author} {\bibfnamefont {J.}~\bibnamefont {Ward}},\
  }\href@noop {} {\bibfield  {journal} {\bibinfo  {journal} {Mol. Phys.},\
  }\textbf {\bibinfo {volume} {20}},\ \bibinfo {pages} {513} (\bibinfo {year}
  {1971})}\BibitemShut {NoStop}%
\bibitem [{\citenamefont {Pavlyukh}\ and\ \citenamefont
  {Berakdar}(2009)}]{pavlyukh_angular_2009}%
  \BibitemOpen
  \bibfield  {author} {\bibinfo {author} {\bibfnamefont {Y.}~\bibnamefont
  {Pavlyukh}}\ and\ \bibinfo {author} {\bibfnamefont {J.}~\bibnamefont
  {Berakdar}},\ }\href@noop {} {\bibfield  {journal} {\bibinfo  {journal}
  {Chem. Phys. Lett.},\ }\textbf {\bibinfo {volume} {468}},\ \bibinfo {pages}
  {313} (\bibinfo {year} {2009})}\BibitemShut {NoStop}%
\bibitem [{\citenamefont {Pavlyukh}\ and\ \citenamefont
  {Berakdar}(2010)}]{pavlyukh_kohn-sham_2010}%
  \BibitemOpen
  \bibfield  {author} {\bibinfo {author} {\bibfnamefont {Y.}~\bibnamefont
  {Pavlyukh}}\ and\ \bibinfo {author} {\bibfnamefont {J.}~\bibnamefont
  {Berakdar}},\ }\href@noop {} {\bibfield  {journal} {\bibinfo  {journal}
  {Phys. Rev. A},\ }\textbf {\bibinfo {volume} {81}},\ \bibinfo {pages}
  {042515} (\bibinfo {year} {2010})}\BibitemShut {NoStop}%
\bibitem [{\citenamefont {Ekardt}(1984)}]{ekardt_dynamical_1984}%
  \BibitemOpen
  \bibfield  {author} {\bibinfo {author} {\bibfnamefont {W.}~\bibnamefont
  {Ekardt}},\ }\href@noop {} {\bibfield  {journal} {\bibinfo  {journal} {Phys.
  Rev. Lett.},\ }\textbf {\bibinfo {volume} {52}},\ \bibinfo {pages} {1925}
  (\bibinfo {year} {1984})}\BibitemShut {NoStop}%
\bibitem [{\citenamefont {Ekardt}(1985)}]{ekardt_size-dependent_1985}%
  \BibitemOpen
  \bibfield  {author} {\bibinfo {author} {\bibfnamefont {W.}~\bibnamefont
  {Ekardt}},\ }\href@noop {} {\bibfield  {journal} {\bibinfo  {journal} {Phys.
  Rev. B},\ }\textbf {\bibinfo {volume} {31}},\ \bibinfo {pages} {6360}
  (\bibinfo {year} {1985})}\BibitemShut {NoStop}%
\bibitem [{\citenamefont {Puska}\ and\ \citenamefont
  {Nieminen}(1993)}]{puska_photoabsorption_1993}%
  \BibitemOpen
  \bibfield  {author} {\bibinfo {author} {\bibfnamefont {M.~J.}\ \bibnamefont
  {Puska}}\ and\ \bibinfo {author} {\bibfnamefont {R.~M.}\ \bibnamefont
  {Nieminen}},\ }\href@noop {} {\bibfield  {journal} {\bibinfo  {journal}
  {Phys. Rev. A},\ }\textbf {\bibinfo {volume} {47}},\ \bibinfo {pages} {1181}
  (\bibinfo {year} {1993})}\BibitemShut {NoStop}%
\bibitem [{\citenamefont {Mukhopadhyay}\ and\ \citenamefont
  {Lundqvist}(1975)}]{mukhopadhyay_density_1975}%
  \BibitemOpen
  \bibfield  {author} {\bibinfo {author} {\bibfnamefont {G.}~\bibnamefont
  {Mukhopadhyay}}\ and\ \bibinfo {author} {\bibfnamefont {S.}~\bibnamefont
  {Lundqvist}},\ }\href@noop {} {\bibfield  {journal} {\bibinfo  {journal}
  {Nuovo Cim. B},\ }\textbf {\bibinfo {volume} {27}},\ \bibinfo {pages} {1}
  (\bibinfo {year} {1975})}\BibitemShut {NoStop}%
\bibitem [{\citenamefont {Vasv\'{a}ri}(1996)}]{vasvari_collective_1996}%
  \BibitemOpen
  \bibfield  {author} {\bibinfo {author} {\bibfnamefont {B.}~\bibnamefont
  {Vasv\'{a}ri}},\ }\href@noop {} {\bibfield  {journal} {\bibinfo  {journal}
  {Z. Phys. B: Condens. Matter},\ }\textbf {\bibinfo {volume} {100}},\ \bibinfo
  {pages} {223} (\bibinfo {year} {1996})}\BibitemShut {NoStop}%
\bibitem [{\citenamefont {Vasv\'{a}ri}(1997)}]{vasvari_collective_1997}%
  \BibitemOpen
  \bibfield  {author} {\bibinfo {author} {\bibfnamefont {B.}~\bibnamefont
  {Vasv\'{a}ri}},\ }\href@noop {} {\bibfield  {journal} {\bibinfo  {journal}
  {Phys. Rev. B},\ }\textbf {\bibinfo {volume} {55}},\ \bibinfo {pages} {7993}
  (\bibinfo {year} {1997})}\BibitemShut {NoStop}%
\bibitem [{\citenamefont {Pavlyukh}\ \emph {et~al.}(2012)\citenamefont
  {Pavlyukh}, \citenamefont {Berakdar},\ and\ \citenamefont
  {K\"{o}ksal}}]{pavlyukh_fast_2012}%
  \BibitemOpen
  \bibfield  {author} {\bibinfo {author} {\bibfnamefont {Y.}~\bibnamefont
  {Pavlyukh}}, \bibinfo {author} {\bibfnamefont {J.}~\bibnamefont {Berakdar}},
  \ and\ \bibinfo {author} {\bibfnamefont {K.}~\bibnamefont {K\"{o}ksal}},\
  }\href@noop {} {\bibfield  {journal} {\bibinfo  {journal} {Phys. Rev. B},\
  }\textbf {\bibinfo {volume} {85}},\ \bibinfo {pages} {195418} (\bibinfo
  {year} {2012})}\BibitemShut {NoStop}%
\bibitem [{\citenamefont {Ichikawa}(2011)}]{ichikawa_theory_2011}%
  \BibitemOpen
  \bibfield  {author} {\bibinfo {author} {\bibfnamefont {M.}~\bibnamefont
  {Ichikawa}},\ }\href@noop {} {\bibfield  {journal} {\bibinfo  {journal} {J.
  Phys. Soc. Jpn.},\ }\textbf {\bibinfo {volume} {80}},\ \bibinfo {pages}
  {044606} (\bibinfo {year} {2011})}\BibitemShut {NoStop}%
\bibitem [{\citenamefont {Wang}\ \emph {et~al.}(1973)\citenamefont {Wang},
  \citenamefont {Chen},\ and\ \citenamefont {Bower}}]{wang_second_1973}%
  \BibitemOpen
  \bibfield  {author} {\bibinfo {author} {\bibfnamefont {C.}~\bibnamefont
  {Wang}}, \bibinfo {author} {\bibfnamefont {J.}~\bibnamefont {Chen}}, \ and\
  \bibinfo {author} {\bibfnamefont {J.}~\bibnamefont {Bower}},\ }\href@noop {}
  {\bibfield  {journal} {\bibinfo  {journal} {Optics Commun.},\ }\textbf
  {\bibinfo {volume} {8}},\ \bibinfo {pages} {275} (\bibinfo {year}
  {1973})}\BibitemShut {NoStop}%
\bibitem [{\citenamefont {Apell}(1983)}]{apell_non-local_1983}%
  \BibitemOpen
  \bibfield  {author} {\bibinfo {author} {\bibfnamefont {P.}~\bibnamefont
  {Apell}},\ }\href@noop {} {\bibfield  {journal} {\bibinfo  {journal} {Phys.
  Scr.},\ }\textbf {\bibinfo {volume} {27}},\ \bibinfo {pages} {211} (\bibinfo
  {year} {1983})}\BibitemShut {NoStop}%
\bibitem [{\citenamefont {Hua}\ and\ \citenamefont
  {Gersten}(1986)}]{hua_theory_1986}%
  \BibitemOpen
  \bibfield  {author} {\bibinfo {author} {\bibfnamefont {X.~M.}\ \bibnamefont
  {Hua}}\ and\ \bibinfo {author} {\bibfnamefont {J.~I.}\ \bibnamefont
  {Gersten}},\ }\href@noop {} {\bibfield  {journal} {\bibinfo  {journal} {Phys.
  Rev. B},\ }\textbf {\bibinfo {volume} {33}},\ \bibinfo {pages} {3756}
  (\bibinfo {year} {1986})}\BibitemShut {NoStop}%
\bibitem [{\citenamefont {Martorell}\ \emph {et~al.}(1997)\citenamefont
  {Martorell}, \citenamefont {Vilaseca},\ and\ \citenamefont
  {Corbal\'{a}n}}]{martorell_scattering_1997}%
  \BibitemOpen
  \bibfield  {author} {\bibinfo {author} {\bibfnamefont {J.}~\bibnamefont
  {Martorell}}, \bibinfo {author} {\bibfnamefont {R.}~\bibnamefont {Vilaseca}},
  \ and\ \bibinfo {author} {\bibfnamefont {R.}~\bibnamefont {Corbal\'{a}n}},\
  }\href@noop {} {\bibfield  {journal} {\bibinfo  {journal} {Phys. Rev. A},\
  }\textbf {\bibinfo {volume} {55}},\ \bibinfo {pages} {4520} (\bibinfo {year}
  {1997})}\BibitemShut {NoStop}%
\bibitem [{\citenamefont {Valencia}\ \emph {et~al.}(2003)\citenamefont
  {Valencia}, \citenamefont {M\'{e}ndez},\ and\ \citenamefont
  {Mendoza}}]{valencia_second-harmonic_2003}%
  \BibitemOpen
  \bibfield  {author} {\bibinfo {author} {\bibfnamefont {C.}~\bibnamefont
  {Valencia}}, \bibinfo {author} {\bibfnamefont {E.}~\bibnamefont
  {M\'{e}ndez}}, \ and\ \bibinfo {author} {\bibfnamefont {B.}~\bibnamefont
  {Mendoza}},\ }\href@noop {} {\bibfield  {journal} {\bibinfo  {journal} {J.
  Opt. Soc. Am. B},\ }\textbf {\bibinfo {volume} {20}},\ \bibinfo {pages}
  {2150} (\bibinfo {year} {2003})}\BibitemShut {NoStop}%
\bibitem [{\citenamefont {Zeng}\ \emph {et~al.}(2009)\citenamefont {Zeng},
  \citenamefont {Hoyer}, \citenamefont {Liu}, \citenamefont {Koch},\ and\
  \citenamefont {Moloney}}]{zeng_classical_2009-1}%
  \BibitemOpen
  \bibfield  {author} {\bibinfo {author} {\bibfnamefont {Y.}~\bibnamefont
  {Zeng}}, \bibinfo {author} {\bibfnamefont {W.}~\bibnamefont {Hoyer}},
  \bibinfo {author} {\bibfnamefont {J.}~\bibnamefont {Liu}}, \bibinfo {author}
  {\bibfnamefont {S.~W.}\ \bibnamefont {Koch}}, \ and\ \bibinfo {author}
  {\bibfnamefont {J.~V.}\ \bibnamefont {Moloney}},\ }\href@noop {} {\bibfield
  {journal} {\bibinfo  {journal} {Phys. Rev. B},\ }\textbf {\bibinfo {volume}
  {79}},\ \bibinfo {pages} {235109} (\bibinfo {year} {2009})}\BibitemShut
  {NoStop}%
\bibitem [{\citenamefont {Scandolo}\ and\ \citenamefont
  {Bassani}(1995)}]{scandolo_kramers-kronig_1995}%
  \BibitemOpen
  \bibfield  {author} {\bibinfo {author} {\bibfnamefont {S.}~\bibnamefont
  {Scandolo}}\ and\ \bibinfo {author} {\bibfnamefont {F.}~\bibnamefont
  {Bassani}},\ }\href@noop {} {\bibfield  {journal} {\bibinfo  {journal} {Phys.
  Rev. B},\ }\textbf {\bibinfo {volume} {51}},\ \bibinfo {pages} {6925}
  (\bibinfo {year} {1995})}\BibitemShut {NoStop}%
\bibitem [{\citenamefont {Satitkovitchai}(2009)}]{satitkovitchai_dynamic_2009}%
  \BibitemOpen
  \bibfield  {author} {\bibinfo {author} {\bibfnamefont {K.}~\bibnamefont
  {Satitkovitchai}},\ }\href@noop {} {\bibfield  {journal} {\bibinfo  {journal}
  {{arXiv:0910.2478}}} (\bibinfo {year} {2009})}\BibitemShut {NoStop}%
\bibitem [{\citenamefont {Ward}(1965)}]{ward_calculation_1965}%
  \BibitemOpen
  \bibfield  {author} {\bibinfo {author} {\bibfnamefont {J.~F.}\ \bibnamefont
  {Ward}},\ }\href@noop {} {\bibfield  {journal} {\bibinfo  {journal} {Rev.
  Mod. Phys.},\ }\textbf {\bibinfo {volume} {37}},\ \bibinfo {pages} {1}
  (\bibinfo {year} {1965})}\BibitemShut {NoStop}%
\bibitem [{\citenamefont {Wang}\ and\ \citenamefont
  {Carter}(2002)}]{wang_orbital-free_2002}%
  \BibitemOpen
  \bibfield  {author} {\bibinfo {author} {\bibfnamefont {Y.}~\bibnamefont
  {Wang}}\ and\ \bibinfo {author} {\bibfnamefont {E.}~\bibnamefont {Carter}},\
  }in\ \href@noop {} {\emph {\bibinfo {booktitle} {Theoretical Methods in
  Condensed Phase Chemistry}}},\ \bibinfo {series and number} {Progress in
  Theoretical Chemistry and Physics}\ (\bibinfo  {publisher} {Kluwer Academic
  Publishers},\ \bibinfo {address} {Dordrecht},\ \bibinfo {year} {2002})\ pp.\
  \bibinfo {pages} {117--184}\BibitemShut {NoStop}%
\bibitem [{\citenamefont {Liebsch}\ and\ \citenamefont
  {Schaich}(1989)}]{liebsch_second-harmonic_1989}%
  \BibitemOpen
  \bibfield  {author} {\bibinfo {author} {\bibfnamefont {A.}~\bibnamefont
  {Liebsch}}\ and\ \bibinfo {author} {\bibfnamefont {W.~L.}\ \bibnamefont
  {Schaich}},\ }\href@noop {} {\bibfield  {journal} {\bibinfo  {journal} {Phys.
  Rev. B},\ }\textbf {\bibinfo {volume} {40}},\ \bibinfo {pages} {5401}
  (\bibinfo {year} {1989})}\BibitemShut {NoStop}%
\bibitem [{\citenamefont {Zangwill}\ and\ \citenamefont
  {Soven}(1980)}]{zangwill_density-functional_1980}%
  \BibitemOpen
  \bibfield  {author} {\bibinfo {author} {\bibfnamefont {A.}~\bibnamefont
  {Zangwill}}\ and\ \bibinfo {author} {\bibfnamefont {P.}~\bibnamefont
  {Soven}},\ }\href@noop {} {\bibfield  {journal} {\bibinfo  {journal} {Phys.
  Rev. A},\ }\textbf {\bibinfo {volume} {21}},\ \bibinfo {pages} {1561}
  (\bibinfo {year} {1980})}\BibitemShut {NoStop}%
\bibitem [{\citenamefont {Jackson}(1999)}]{jackson_classical_1999}%
  \BibitemOpen
  \bibfield  {author} {\bibinfo {author} {\bibfnamefont {J.}~\bibnamefont
  {Jackson}},\ }\href@noop {} {\emph {\bibinfo {title} {Classical
  electrodynamics}}},\ \bibinfo {edition} {3rd}\ ed.\ (\bibinfo  {publisher}
  {Wiley},\ \bibinfo {address} {New York},\ \bibinfo {year} {1999})\BibitemShut
  {NoStop}%
\bibitem [{\citenamefont {Rudnick}\ and\ \citenamefont
  {Stern}(1971)}]{rudnick_second-harmonic_1971}%
  \BibitemOpen
  \bibfield  {author} {\bibinfo {author} {\bibfnamefont {J.}~\bibnamefont
  {Rudnick}}\ and\ \bibinfo {author} {\bibfnamefont {E.~A.}\ \bibnamefont
  {Stern}},\ }\href@noop {} {\bibfield  {journal} {\bibinfo  {journal} {Phys.
  Rev. B},\ }\textbf {\bibinfo {volume} {4}},\ \bibinfo {pages} {4274}
  (\bibinfo {year} {1971})}\BibitemShut {NoStop}%
\end{thebibliography}
\end{document}